\documentclass[orivec, envcountsame]{llncs}
\usepackage[english]{babel}

\usepackage{amssymb, amsmath}
\usepackage[backend=biber]{biblatex}
\usepackage{csquotes}
\usepackage{graphicx}
\graphicspath{{images/}}
\usepackage[colorlinks=true, allcolors=blue]{hyperref}

\usepackage{tikz}
\usetikzlibrary{automata, positioning, arrows, arrows.meta, shapes, shapes.geometric, fit, calc}
 \newcommand{\eq}{=} %

\usepackage{isabelle}
\usepackage{isabellesym}

\usepackage{ifthen}
\usepackage{bbold}

\usepackage{hyperref}

\usepackage{subfiles}

\newcommand{\isalogo}{\includegraphics[width=9pt]{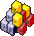}}
\newcommand{\isalink}[1]{\href{#1}{\isalogo}}

\definecolor{IsaBlue}{cmyk}{1.0,0.33,0,.4}
\definecolor{IsaGreen}{cmyk}{1.0,0,1.0,.2}
\definecolor{IsaRed}{cmyk}{0,0.69,0.69,0}
\definecolor{IsaDarkRed}{cmyk}{0,1.0,1.0,0.15}

\isabellestylett

\renewcommand{\isastyle}{\isastyleminor}

\newcommand{\isakw}[1]{{\bfseries\ttfamily\def\isachardot{.}\def\isacharunderscore{\isacharunderscorekeyword}%
\def\isacharbraceleft{\{}\def\isacharbraceright{\}}#1}}

\renewcommand{\isacommand}[1]{\textcolor{IsaBlue}{\isakw{#1}}}
\renewcommand{\isakeyword}[1]{\textcolor{IsaGreen}{\isakw{#1}}}

\newcommand{\DefineSnippet}[2]{%
\expandafter\newcommand\csname snippet--#1\endcsname{%
 \begin{quote}
 \begin{isabelle}
 #2
 \end{isabelle}
 \end{quote}}}
\newcommand{\Snippet}[1]{%
\ifcsname snippet--#1\endcsname{\csname snippet--#1\endcsname}%
\else+++++++ERROR: Snippet ``#1 not defined+++++++ \fi}

\DefineSnippet{Wiener_kernel}{
\isacommand{definition}\isamarkupfalse%
\ Wiener{\isacharunderscore}{\kern0pt}kernel\ {\isacharcolon}{\kern0pt}{\isacharcolon}{\kern0pt}\ {\isachardoublequoteopen}real\ {\isasymRightarrow}\ real\ hkernel{\isachardoublequoteclose}\ \isakeyword{where}\isanewline
{\isachardoublequoteopen}Wiener{\isacharunderscore}{\kern0pt}kernel\ i\ {\isasymequiv}\ if\ i\ {\isacharequal}{\kern0pt}\ {\isadigit{0}}\ then\ hkernel\ {\isacharparenleft}{\kern0pt}return{\isacharunderscore}{\kern0pt}kernel\ borel{\isacharparenright}{\kern0pt}\isanewline
\ \ \ else\ hkernel{\isacharunderscore}{\kern0pt}of\ borel\ {\isacharparenleft}{\kern0pt}{\isasymlambda}x\ dy{\isachardot}{\kern0pt}\ {\isacharparenleft}{\kern0pt}return\ borel\ x\ {\isasymstar}\ density\ lborel\ {\isacharparenleft}{\kern0pt}normal{\isacharunderscore}{\kern0pt}density\ {\isadigit{0}}\ i{\isacharparenright}{\kern0pt}{\isacharparenright}{\kern0pt}\ dy{\isacharparenright}{\kern0pt}{\isachardoublequoteclose}%
}%
\DefineSnippet{brownian_motion}{
\isacommand{locale}\isamarkupfalse%
\ brownian{\isacharunderscore}{\kern0pt}motion\ {\isacharequal}{\kern0pt}\isanewline
\isakeyword{fixes}\ W\ {\isacharcolon}{\kern0pt}{\isacharcolon}{\kern0pt}\ {\isachardoublequoteopen}{\isacharparenleft}{\kern0pt}real{\isacharcomma}{\kern0pt}\ {\isacharprime}{\kern0pt}a{\isacharcomma}{\kern0pt}\ real{\isacharparenright}{\kern0pt}\ stochastic{\isacharunderscore}{\kern0pt}process{\isachardoublequoteclose}\isanewline
\isakeyword{assumes}\ init{\isacharunderscore}{\kern0pt}{\isadigit{0}}{\isacharbrackleft}{\kern0pt}simp{\isacharbrackright}{\kern0pt}{\isacharcolon}{\kern0pt}\ {\isachardoublequoteopen}{\isasymP}{\isacharparenleft}{\kern0pt}x\ in\ proc{\isacharunderscore}{\kern0pt}source\ W{\isachardot}{\kern0pt}\ W\ {\isadigit{0}}\ x\ {\isacharequal}{\kern0pt}\ {\isadigit{0}}{\isacharparenright}{\kern0pt}\ {\isacharequal}{\kern0pt}\ {\isadigit{1}}{\isachardoublequoteclose}\isanewline
\ \ \ \ \isakeyword{and}\ indep{\isacharunderscore}{\kern0pt}increments{\isacharcolon}{\kern0pt}\ {\isachardoublequoteopen}indep{\isacharunderscore}{\kern0pt}increments\ W{\isachardoublequoteclose}\isanewline
\ \ \ \ \isakeyword{and}\ normal{\isacharunderscore}{\kern0pt}increments{\isacharcolon}{\kern0pt}\ {\isachardoublequoteopen}{\isasymAnd}s\ t{\isachardot}{\kern0pt}\ {\isadigit{0}}\ {\isasymle}\ s\ {\isasymand}\ s\ {\isacharless}{\kern0pt}\ t\ {\isasymLongrightarrow}\isanewline
\ distributed\ {\isacharparenleft}{\kern0pt}proc{\isacharunderscore}{\kern0pt}source\ W{\isacharparenright}{\kern0pt}\ borel\ {\isacharparenleft}{\kern0pt}{\isasymlambda}v{\isachardot}{\kern0pt}\ W\ t\ v\ {\isacharminus}{\kern0pt}\ W\ s\ v{\isacharparenright}{\kern0pt}\isanewline
\ \ \ \ \ \ \ \ \ \ \ \ \ {\isacharparenleft}{\kern0pt}normal{\isacharunderscore}{\kern0pt}density\ {\isadigit{0}}\ {\isacharparenleft}{\kern0pt}sqrt\ {\isacharparenleft}{\kern0pt}t{\isacharminus}{\kern0pt}s{\isacharparenright}{\kern0pt}{\isacharparenright}{\kern0pt}{\isacharparenright}{\kern0pt}{\isachardoublequoteclose}\isanewline
\ \ \ \ \isakeyword{and}\ AE{\isacharunderscore}{\kern0pt}continuous{\isacharcolon}{\kern0pt}\ {\isachardoublequoteopen}AE\ x\ in\ proc{\isacharunderscore}{\kern0pt}source\ W{\isachardot}{\kern0pt}\ continuous{\isacharunderscore}{\kern0pt}on\ {\isacharbraceleft}{\kern0pt}{\isadigit{0}}{\isachardot}{\kern0pt}{\isachardot}{\kern0pt}{\isacharbraceright}{\kern0pt}\ {\isacharparenleft}{\kern0pt}{\isasymlambda}t{\isachardot}{\kern0pt}\ W\ t\ x{\isacharparenright}{\kern0pt}{\isachardoublequoteclose}%
}%
\DefineSnippet{kernel_product}{
\isacommand{definition}\isamarkupfalse%
\ kernel{\isacharunderscore}{\kern0pt}product\ {\isacharcolon}{\kern0pt}{\isacharcolon}{\kern0pt}\ {\isachardoublequoteopen}{\isacharparenleft}{\kern0pt}{\isacharprime}{\kern0pt}a{\isacharcomma}{\kern0pt}\ {\isacharprime}{\kern0pt}b{\isacharparenright}{\kern0pt}\ kernel\ {\isasymRightarrow}\ {\isacharparenleft}{\kern0pt}{\isacharprime}{\kern0pt}a\ {\isasymtimes}\ {\isacharprime}{\kern0pt}b{\isacharcomma}{\kern0pt}\ {\isacharprime}{\kern0pt}c{\isacharparenright}{\kern0pt}\ kernel\ {\isasymRightarrow}\ {\isacharparenleft}{\kern0pt}{\isacharprime}{\kern0pt}a{\isacharcomma}{\kern0pt}\ {\isacharprime}{\kern0pt}b\ {\isasymtimes}\ {\isacharprime}{\kern0pt}c{\isacharparenright}{\kern0pt}\ kernel{\isachardoublequoteclose}\ {\isacharparenleft}{\kern0pt}\isakeyword{infixr}\ {\isachardoublequoteopen}{\isacharparenleft}{\kern0pt}{\isasymOtimes}\isactrlsub K{\isacharparenright}{\kern0pt}{\isachardoublequoteclose}\ {\isadigit{9}}{\isadigit{0}}{\isacharparenright}{\kern0pt}\ \isakeyword{where}\isanewline
{\isachardoublequoteopen}kernel{\isacharunderscore}{\kern0pt}product\ K{\isacharunderscore}{\kern0pt}{\isadigit{1}}\ K{\isacharunderscore}{\kern0pt}{\isadigit{2}}\ {\isasymequiv}\ kernel{\isacharunderscore}{\kern0pt}of\ {\isacharparenleft}{\kern0pt}kernel{\isacharunderscore}{\kern0pt}source\ K{\isacharunderscore}{\kern0pt}{\isadigit{1}}{\isacharparenright}{\kern0pt}\ {\isacharparenleft}{\kern0pt}kernel{\isacharunderscore}{\kern0pt}target\ K{\isacharunderscore}{\kern0pt}{\isadigit{1}}\ {\isasymOtimes}\isactrlsub M\ kernel{\isacharunderscore}{\kern0pt}target\ K{\isacharunderscore}{\kern0pt}{\isadigit{2}}{\isacharparenright}{\kern0pt}\isanewline
\ \ {\isacharparenleft}{\kern0pt}{\isasymlambda}{\isasymomega}\isactrlsub {\isadigit{0}}\ A{\isachardot}{\kern0pt}\ {\isasymintegral}\isactrlsup {\isacharplus}{\kern0pt}{\isasymomega}\isactrlsub {\isadigit{1}}{\isachardot}{\kern0pt}\ {\isacharparenleft}{\kern0pt}{\isasymintegral}\isactrlsup {\isacharplus}{\kern0pt}{\isasymomega}\isactrlsub {\isadigit{2}}{\isachardot}{\kern0pt}\ indicator\ A\ {\isacharparenleft}{\kern0pt}{\isasymomega}\isactrlsub {\isadigit{1}}{\isacharcomma}{\kern0pt}\ {\isasymomega}\isactrlsub {\isadigit{2}}{\isacharparenright}{\kern0pt}{\isasympartial}kernel{\isacharunderscore}{\kern0pt}measure\ K{\isacharunderscore}{\kern0pt}{\isadigit{2}}\ {\isacharparenleft}{\kern0pt}{\isasymomega}\isactrlsub {\isadigit{0}}{\isacharcomma}{\kern0pt}\ {\isasymomega}\isactrlsub {\isadigit{1}}{\isacharparenright}{\kern0pt}{\isacharparenright}{\kern0pt}{\isasympartial}kernel{\isacharunderscore}{\kern0pt}measure\ K{\isacharunderscore}{\kern0pt}{\isadigit{1}}\ {\isasymomega}\isactrlsub {\isadigit{0}}{\isacharparenright}{\kern0pt}{\isachardoublequoteclose}%
}%
\DefineSnippet{kernel_product_partial}{
\isacommand{definition}\isamarkupfalse%
\ kernel{\isacharunderscore}{\kern0pt}product{\isacharunderscore}{\kern0pt}partial\ {\isacharcolon}{\kern0pt}{\isacharcolon}{\kern0pt}\ {\isachardoublequoteopen}{\isacharparenleft}{\kern0pt}{\isacharprime}{\kern0pt}a{\isacharcomma}{\kern0pt}\ {\isacharprime}{\kern0pt}b{\isacharparenright}{\kern0pt}\ kernel\ {\isasymRightarrow}\ {\isacharparenleft}{\kern0pt}{\isacharprime}{\kern0pt}b{\isacharcomma}{\kern0pt}\ {\isacharprime}{\kern0pt}c{\isacharparenright}{\kern0pt}\ kernel\ {\isasymRightarrow}\ {\isacharparenleft}{\kern0pt}{\isacharprime}{\kern0pt}a{\isacharcomma}{\kern0pt}\ {\isacharprime}{\kern0pt}b\ {\isasymtimes}\ {\isacharprime}{\kern0pt}c{\isacharparenright}{\kern0pt}\ kernel{\isachardoublequoteclose}\ {\isacharparenleft}{\kern0pt}\isakeyword{infixr}\ {\isachardoublequoteopen}{\isacharparenleft}{\kern0pt}{\isasymOtimes}\isactrlsub P{\isacharparenright}{\kern0pt}{\isachardoublequoteclose}\ {\isadigit{9}}{\isadigit{0}}{\isacharparenright}{\kern0pt}\ \isakeyword{where}\isanewline
{\isachardoublequoteopen}kernel{\isacharunderscore}{\kern0pt}product{\isacharunderscore}{\kern0pt}partial\ K{\isacharunderscore}{\kern0pt}{\isadigit{1}}\ K{\isacharunderscore}{\kern0pt}{\isadigit{2}}\ {\isasymequiv}\ K{\isacharunderscore}{\kern0pt}{\isadigit{1}}\ {\isasymOtimes}\isactrlsub K\ {\isacharparenleft}{\kern0pt}kernel{\isacharunderscore}{\kern0pt}of\ {\isacharparenleft}{\kern0pt}kernel{\isacharunderscore}{\kern0pt}source\ K{\isacharunderscore}{\kern0pt}{\isadigit{1}}\ {\isasymOtimes}\isactrlsub M\ kernel{\isacharunderscore}{\kern0pt}source\ K{\isacharunderscore}{\kern0pt}{\isadigit{2}}{\isacharparenright}{\kern0pt}\ {\isacharparenleft}{\kern0pt}kernel{\isacharunderscore}{\kern0pt}target\ K{\isacharunderscore}{\kern0pt}{\isadigit{2}}{\isacharparenright}{\kern0pt}\isanewline
\ {\isacharparenleft}{\kern0pt}{\isasymlambda}{\isasymomega}\ A\isactrlsub {\isadigit{2}}{\isachardot}{\kern0pt}\ kernel\ K{\isacharunderscore}{\kern0pt}{\isadigit{2}}\ {\isacharparenleft}{\kern0pt}snd\ {\isasymomega}{\isacharparenright}{\kern0pt}\ A\isactrlsub {\isadigit{2}}{\isacharparenright}{\kern0pt}{\isacharparenright}{\kern0pt}{\isachardoublequoteclose}%
}%
\DefineSnippet{kernel_snd_measurable}{
\isacommand{lemma}\isamarkupfalse%
\ kernel{\isacharunderscore}{\kern0pt}snd{\isacharunderscore}{\kern0pt}measurable{\isacharcolon}{\kern0pt}\isanewline
\ \ \isakeyword{fixes}\ K\ {\isacharcolon}{\kern0pt}{\isacharcolon}{\kern0pt}\ {\isachardoublequoteopen}{\isacharparenleft}{\kern0pt}{\isacharprime}{\kern0pt}a{\isasymtimes}{\isacharprime}{\kern0pt}b{\isacharcomma}{\kern0pt}\ {\isacharprime}{\kern0pt}c{\isacharparenright}{\kern0pt}\ kernel{\isachardoublequoteclose}\isanewline
\ \ \isakeyword{assumes}\ A{\isacharunderscore}{\kern0pt}sets{\isacharcolon}{\kern0pt}\ {\isachardoublequoteopen}A\ {\isasymin}\ sets\ {\isacharparenleft}{\kern0pt}M{\isadigit{1}}\ {\isasymOtimes}\isactrlsub M\ kernel{\isacharunderscore}{\kern0pt}target\ K{\isacharparenright}{\kern0pt}{\isachardoublequoteclose}\isanewline
\ \ \isakeyword{and}\ sets{\isacharunderscore}{\kern0pt}eq{\isacharcolon}{\kern0pt}\ {\isachardoublequoteopen}sets\ {\isacharparenleft}{\kern0pt}kernel{\isacharunderscore}{\kern0pt}source\ K{\isacharparenright}{\kern0pt}\ {\isacharequal}{\kern0pt}\ sets\ {\isacharparenleft}{\kern0pt}M{\isadigit{0}}\ {\isasymOtimes}\isactrlsub M\ M{\isadigit{1}}{\isacharparenright}{\kern0pt}{\isachardoublequoteclose}\isanewline
\ \ \isakeyword{and}\ {\isasymomega}\isactrlsub {\isadigit{0}}{\isacharcolon}{\kern0pt}\ {\isachardoublequoteopen}{\isasymomega}\isactrlsub {\isadigit{0}}\ {\isasymin}\ space\ M{\isadigit{0}}{\isachardoublequoteclose}\isanewline
\ \ \isakeyword{and}\ finite{\isacharunderscore}{\kern0pt}kernel{\isacharcolon}{\kern0pt}\ {\isachardoublequoteopen}finite{\isacharunderscore}{\kern0pt}kernel\ K{\isachardoublequoteclose}\isanewline
\ \ \isakeyword{shows}\ {\isachardoublequoteopen}{\isacharparenleft}{\kern0pt}{\isasymlambda}w{\isachardot}{\kern0pt}\ kernel\ K\ {\isacharparenleft}{\kern0pt}{\isasymomega}\isactrlsub {\isadigit{0}}{\isacharcomma}{\kern0pt}\ w{\isacharparenright}{\kern0pt}\ {\isacharbraceleft}{\kern0pt}{\isasymomega}\isactrlsub {\isadigit{2}}{\isachardot}{\kern0pt}\ {\isacharparenleft}{\kern0pt}w{\isacharcomma}{\kern0pt}\ {\isasymomega}\isactrlsub {\isadigit{2}}{\isacharparenright}{\kern0pt}\ {\isasymin}\ A{\isacharbraceright}{\kern0pt}{\isacharparenright}{\kern0pt}\ {\isasymin}\ borel{\isacharunderscore}{\kern0pt}measurable\ M{\isadigit{1}}{\isachardoublequoteclose}%
}%
\DefineSnippet{kernel_product_transition_kernel}{
\isacommand{theorem}\isamarkupfalse%
\ kernel{\isacharunderscore}{\kern0pt}product{\isacharunderscore}{\kern0pt}transition{\isacharunderscore}{\kern0pt}kernel{\isacharcolon}{\kern0pt}\isanewline
\ \ \isakeyword{fixes}\ K{\isacharunderscore}{\kern0pt}{\isadigit{1}}\ {\isacharcolon}{\kern0pt}{\isacharcolon}{\kern0pt}\ {\isachardoublequoteopen}{\isacharparenleft}{\kern0pt}{\isacharprime}{\kern0pt}a{\isacharcomma}{\kern0pt}\ {\isacharprime}{\kern0pt}b{\isacharparenright}{\kern0pt}\ kernel{\isachardoublequoteclose}\isanewline
\ \ \ \ \isakeyword{and}\ K{\isacharunderscore}{\kern0pt}{\isadigit{2}}\ {\isacharcolon}{\kern0pt}{\isacharcolon}{\kern0pt}\ {\isachardoublequoteopen}{\isacharparenleft}{\kern0pt}{\isacharprime}{\kern0pt}a{\isasymtimes}{\isacharprime}{\kern0pt}b{\isacharcomma}{\kern0pt}\ {\isacharprime}{\kern0pt}c{\isacharparenright}{\kern0pt}\ kernel{\isachardoublequoteclose}\isanewline
\ \ \isakeyword{assumes}\ finite{\isacharcolon}{\kern0pt}\ {\isachardoublequoteopen}finite{\isacharunderscore}{\kern0pt}kernel\ K{\isacharunderscore}{\kern0pt}{\isadigit{1}}{\isachardoublequoteclose}\ {\isachardoublequoteopen}finite{\isacharunderscore}{\kern0pt}kernel\ K{\isacharunderscore}{\kern0pt}{\isadigit{2}}{\isachardoublequoteclose}\isanewline
\ \ \ \ \ \ \isakeyword{and}\ eq{\isacharcolon}{\kern0pt}\ {\isachardoublequoteopen}sets\ {\isacharparenleft}{\kern0pt}kernel{\isacharunderscore}{\kern0pt}source\ K{\isacharunderscore}{\kern0pt}{\isadigit{2}}{\isacharparenright}{\kern0pt}\ {\isacharequal}{\kern0pt}\ sets\ {\isacharparenleft}{\kern0pt}kernel{\isacharunderscore}{\kern0pt}source\ K{\isacharunderscore}{\kern0pt}{\isadigit{1}}\ {\isasymOtimes}\isactrlsub M\ kernel{\isacharunderscore}{\kern0pt}target\ K{\isacharunderscore}{\kern0pt}{\isadigit{1}}{\isacharparenright}{\kern0pt}{\isachardoublequoteclose}\isanewline
\ \ \ \ \isakeyword{shows}\ {\isachardoublequoteopen}transition{\isacharunderscore}{\kern0pt}kernel\ {\isacharparenleft}{\kern0pt}kernel{\isacharunderscore}{\kern0pt}source\ K{\isacharunderscore}{\kern0pt}{\isadigit{1}}{\isacharparenright}{\kern0pt}\ {\isacharparenleft}{\kern0pt}kernel{\isacharunderscore}{\kern0pt}target\ K{\isacharunderscore}{\kern0pt}{\isadigit{1}}\ {\isasymOtimes}\isactrlsub M\ kernel{\isacharunderscore}{\kern0pt}target\ K{\isacharunderscore}{\kern0pt}{\isadigit{2}}{\isacharparenright}{\kern0pt}\isanewline
\ \ \ \ {\isacharparenleft}{\kern0pt}{\isasymlambda}{\isasymomega}\isactrlsub {\isadigit{0}}\ A{\isachardot}{\kern0pt}\ {\isasymintegral}\isactrlsup {\isacharplus}{\kern0pt}{\isasymomega}\isactrlsub {\isadigit{1}}{\isachardot}{\kern0pt}\ {\isacharparenleft}{\kern0pt}{\isasymintegral}\isactrlsup {\isacharplus}{\kern0pt}{\isasymomega}\isactrlsub {\isadigit{2}}{\isachardot}{\kern0pt}\ indicator\ A\ {\isacharparenleft}{\kern0pt}{\isasymomega}\isactrlsub {\isadigit{1}}{\isacharcomma}{\kern0pt}\ {\isasymomega}\isactrlsub {\isadigit{2}}{\isacharparenright}{\kern0pt}\ {\isasympartial}kernel{\isacharunderscore}{\kern0pt}measure\ K{\isacharunderscore}{\kern0pt}{\isadigit{2}}\ {\isacharparenleft}{\kern0pt}{\isasymomega}\isactrlsub {\isadigit{0}}{\isacharcomma}{\kern0pt}\ {\isasymomega}\isactrlsub {\isadigit{1}}{\isacharparenright}{\kern0pt}{\isacharparenright}{\kern0pt}\ {\isasympartial}kernel{\isacharunderscore}{\kern0pt}measure\ K{\isacharunderscore}{\kern0pt}{\isadigit{1}}\ {\isasymomega}\isactrlsub {\isadigit{0}}{\isacharparenright}{\kern0pt}{\isachardoublequoteclose}\isanewline
\ \ {\isacharparenleft}{\kern0pt}\isakeyword{is}\ {\isachardoublequoteopen}transition{\isacharunderscore}{\kern0pt}kernel\ {\isacharquery}{\kern0pt}{\isasymOmega}\isactrlsub {\isadigit{1}}\ {\isacharquery}{\kern0pt}{\isasymOmega}\isactrlsub {\isadigit{2}}\ {\isacharquery}{\kern0pt}{\isasymkappa}{\isachardoublequoteclose}{\isacharparenright}{\kern0pt}%
}%
\DefineSnippet{kernel_semidirect_product}{
\isacommand{definition}\isamarkupfalse%
\ kernel{\isacharunderscore}{\kern0pt}semidirect{\isacharunderscore}{\kern0pt}product\ {\isacharcolon}{\kern0pt}{\isacharcolon}{\kern0pt}\ {\isachardoublequoteopen}{\isacharprime}{\kern0pt}a\ measure\ {\isasymRightarrow}\ {\isacharparenleft}{\kern0pt}{\isacharprime}{\kern0pt}a{\isacharcomma}{\kern0pt}\ {\isacharprime}{\kern0pt}b{\isacharparenright}{\kern0pt}\ kernel\ {\isasymRightarrow}\ {\isacharparenleft}{\kern0pt}{\isacharprime}{\kern0pt}a\ {\isasymtimes}\ {\isacharprime}{\kern0pt}b{\isacharparenright}{\kern0pt}\ measure{\isachardoublequoteclose}\ {\isacharparenleft}{\kern0pt}\isakeyword{infixr}\ {\isachardoublequoteopen}{\isacharparenleft}{\kern0pt}{\isasymOtimes}\isactrlsub S{\isacharparenright}{\kern0pt}{\isachardoublequoteclose}\ {\isadigit{7}}{\isadigit{0}}{\isacharparenright}{\kern0pt}\isanewline
\ \ \isakeyword{where}\ {\isachardoublequoteopen}M\ {\isasymOtimes}\isactrlsub S\ K\ {\isacharequal}{\kern0pt}\ kernel{\isacharunderscore}{\kern0pt}measure\ {\isacharparenleft}{\kern0pt}emeasure{\isacharunderscore}{\kern0pt}kernel\ M\ M\ {\isasymOtimes}\isactrlsub P\ K{\isacharparenright}{\kern0pt}\ {\isacharparenleft}{\kern0pt}SOME\ {\isasymomega}{\isachardot}{\kern0pt}\ {\isasymomega}\ {\isasymin}\ space\ {\isacharparenleft}{\kern0pt}kernel{\isacharunderscore}{\kern0pt}source\ K{\isacharparenright}{\kern0pt}{\isacharparenright}{\kern0pt}{\isachardoublequoteclose}%
}%
\DefineSnippet{emeasure_semidirect_product}{
\isacommand{lemma}\isamarkupfalse%
\ emeasure{\isacharunderscore}{\kern0pt}semidirect{\isacharunderscore}{\kern0pt}product{\isacharcolon}{\kern0pt}\isanewline
\ \ \isakeyword{assumes}\ {\isachardoublequoteopen}A\ {\isasymin}\ sets\ {\isacharparenleft}{\kern0pt}M\ {\isasymOtimes}\isactrlsub S\ K{\isacharparenright}{\kern0pt}{\isachardoublequoteclose}\isanewline
\ \ \isakeyword{shows}\ {\isachardoublequoteopen}emeasure\ {\isacharparenleft}{\kern0pt}M\ {\isasymOtimes}\isactrlsub S\ K{\isacharparenright}{\kern0pt}\ A\ {\isacharequal}{\kern0pt}\ {\isasymintegral}\isactrlsup {\isacharplus}{\kern0pt}\ {\isasymomega}\isactrlsub {\isadigit{1}}{\isachardot}{\kern0pt}\ {\isasymintegral}\isactrlsup {\isacharplus}{\kern0pt}\ {\isasymomega}\isactrlsub {\isadigit{2}}{\isachardot}{\kern0pt}\ indicator\ A\ {\isacharparenleft}{\kern0pt}{\isasymomega}\isactrlsub {\isadigit{1}}{\isacharcomma}{\kern0pt}\ {\isasymomega}\isactrlsub {\isadigit{2}}{\isacharparenright}{\kern0pt}\ {\isasympartial}kernel{\isacharunderscore}{\kern0pt}measure\ K\ {\isasymomega}\isactrlsub {\isadigit{1}}\ {\isasympartial}M{\isachardoublequoteclose}%
}%
\DefineSnippet{kernel_Fubini}{
\isacommand{lemma}\isamarkupfalse%
\ kernel{\isacharunderscore}{\kern0pt}Fubini{\isacharcolon}{\kern0pt}\isanewline
\ \ \isakeyword{assumes}\ f{\isacharbrackleft}{\kern0pt}measurable{\isacharbrackright}{\kern0pt}{\isacharcolon}{\kern0pt}\ {\isachardoublequoteopen}f\ {\isasymin}\ borel{\isacharunderscore}{\kern0pt}measurable\ {\isacharparenleft}{\kern0pt}M\ {\isasymOtimes}\isactrlsub M\ {\isacharparenleft}{\kern0pt}kernel{\isacharunderscore}{\kern0pt}target\ K{\isacharparenright}{\kern0pt}{\isacharparenright}{\kern0pt}{\isachardoublequoteclose}\isanewline
\ \ \isakeyword{shows}\ {\isachardoublequoteopen}{\isacharparenleft}{\kern0pt}{\isasymintegral}\isactrlsup {\isacharplus}{\kern0pt}{\isasymomega}{\isachardot}{\kern0pt}\ f\ {\isasymomega}\ {\isasympartial}{\isacharparenleft}{\kern0pt}M\ {\isasymOtimes}\isactrlsub S\ K{\isacharparenright}{\kern0pt}{\isacharparenright}{\kern0pt}\ {\isacharequal}{\kern0pt}\ {\isacharparenleft}{\kern0pt}{\isasymintegral}\isactrlsup {\isacharplus}{\kern0pt}{\isasymomega}\isactrlsub {\isadigit{1}}{\isachardot}{\kern0pt}\ {\isacharparenleft}{\kern0pt}{\isasymintegral}\isactrlsup {\isacharplus}{\kern0pt}{\isasymomega}\isactrlsub {\isadigit{2}}{\isachardot}{\kern0pt}\ f\ {\isacharparenleft}{\kern0pt}{\isasymomega}\isactrlsub {\isadigit{1}}{\isacharcomma}{\kern0pt}\ {\isasymomega}\isactrlsub {\isadigit{2}}{\isacharparenright}{\kern0pt}\ {\isasympartial}kernel{\isacharunderscore}{\kern0pt}measure\ K\ {\isasymomega}\isactrlsub {\isadigit{1}}{\isacharparenright}{\kern0pt}\ {\isasympartial}M{\isacharparenright}{\kern0pt}{\isachardoublequoteclose}%
}%
\DefineSnippet{PiK}{
\isacommand{primrec}\isamarkupfalse%
\ PiK\ {\isacharcolon}{\kern0pt}{\isacharcolon}{\kern0pt}\ {\isachardoublequoteopen}nat\ {\isasymRightarrow}\ {\isacharparenleft}{\kern0pt}nat\ {\isasymRightarrow}\ {\isacharparenleft}{\kern0pt}{\isacharprime}{\kern0pt}a{\isacharcomma}{\kern0pt}\ {\isacharprime}{\kern0pt}a{\isacharparenright}{\kern0pt}\ kernel{\isacharparenright}{\kern0pt}\ {\isasymRightarrow}\ {\isacharparenleft}{\kern0pt}{\isacharprime}{\kern0pt}a{\isacharcomma}{\kern0pt}\ {\isacharparenleft}{\kern0pt}nat\ {\isasymRightarrow}\ {\isacharprime}{\kern0pt}a{\isacharparenright}{\kern0pt}{\isacharparenright}{\kern0pt}\ kernel{\isachardoublequoteclose}\ \isakeyword{where}\isanewline
{\isachardoublequoteopen}PiK\ {\isadigit{0}}\ K\ {\isacharequal}{\kern0pt}\ kernel{\isacharunderscore}{\kern0pt}of\ {\isacharparenleft}{\kern0pt}kernel{\isacharunderscore}{\kern0pt}source\ {\isacharparenleft}{\kern0pt}K\ {\isadigit{0}}{\isacharparenright}{\kern0pt}{\isacharparenright}{\kern0pt}\ {\isacharparenleft}{\kern0pt}PiM\ {\isacharbraceleft}{\kern0pt}{\isadigit{0}}{\isacharbraceright}{\kern0pt}\ {\isacharparenleft}{\kern0pt}{\isasymlambda}i{\isachardot}{\kern0pt}\ kernel{\isacharunderscore}{\kern0pt}target\ {\isacharparenleft}{\kern0pt}K\ i{\isacharparenright}{\kern0pt}{\isacharparenright}{\kern0pt}{\isacharparenright}{\kern0pt}\isanewline
\ \ \ {\isacharparenleft}{\kern0pt}{\isasymlambda}\ {\isasymomega}\ A{\isacharprime}{\kern0pt}{\isachardot}{\kern0pt}\ {\isacharparenleft}{\kern0pt}K\ {\isadigit{0}}{\isacharparenright}{\kern0pt}\ {\isasymomega}\ {\isacharparenleft}{\kern0pt}{\isacharparenleft}{\kern0pt}{\isasymlambda}x{\isachardot}{\kern0pt}{\isacharparenleft}{\kern0pt}{\isasymlambda}n{\isasymin}{\isacharbraceleft}{\kern0pt}{\isadigit{0}}{\isacharbraceright}{\kern0pt}{\isachardot}{\kern0pt}\ x{\isacharparenright}{\kern0pt}{\isacharparenright}{\kern0pt}\ {\isacharminus}{\kern0pt}{\isacharbackquote}{\kern0pt}\ A{\isacharprime}{\kern0pt}\ {\isasyminter}\ space\ {\isacharparenleft}{\kern0pt}kernel{\isacharunderscore}{\kern0pt}target\ {\isacharparenleft}{\kern0pt}K\ {\isadigit{0}}{\isacharparenright}{\kern0pt}{\isacharparenright}{\kern0pt}{\isacharparenright}{\kern0pt}{\isacharparenright}{\kern0pt}{\isachardoublequoteclose}\ {\isacharbar}{\kern0pt}\isanewline
{\isachardoublequoteopen}PiK\ {\isacharparenleft}{\kern0pt}Suc\ n{\isacharparenright}{\kern0pt}\ K\ {\isacharequal}{\kern0pt}\ kernel{\isacharunderscore}{\kern0pt}of\ {\isacharparenleft}{\kern0pt}kernel{\isacharunderscore}{\kern0pt}source\ {\isacharparenleft}{\kern0pt}K\ {\isadigit{0}}{\isacharparenright}{\kern0pt}{\isacharparenright}{\kern0pt}\ {\isacharparenleft}{\kern0pt}PiM\ {\isacharbraceleft}{\kern0pt}{\isadigit{0}}{\isachardot}{\kern0pt}{\isachardot}{\kern0pt}Suc\ n{\isacharbraceright}{\kern0pt}\ {\isacharparenleft}{\kern0pt}{\isasymlambda}i{\isachardot}{\kern0pt}\ kernel{\isacharunderscore}{\kern0pt}target\ {\isacharparenleft}{\kern0pt}K\ i{\isacharparenright}{\kern0pt}{\isacharparenright}{\kern0pt}{\isacharparenright}{\kern0pt}\isanewline
\ \ {\isacharparenleft}{\kern0pt}{\isasymlambda}{\isasymomega}\ A{\isacharprime}{\kern0pt}{\isachardot}{\kern0pt}\ {\isasymintegral}\isactrlsup {\isacharplus}{\kern0pt}{\isasymomega}\isactrlsub f{\isachardot}{\kern0pt}\ {\isacharparenleft}{\kern0pt}{\isasymintegral}\isactrlsup {\isacharplus}{\kern0pt}{\isasymomega}{\isachardot}{\kern0pt}\ indicator\ A{\isacharprime}{\kern0pt}\ {\isacharparenleft}{\kern0pt}{\isasymomega}\isactrlsub f\ {\isacharparenleft}{\kern0pt}Suc\ n{\isacharcolon}{\kern0pt}{\isacharequal}{\kern0pt}{\isasymomega}{\isacharparenright}{\kern0pt}{\isacharparenright}{\kern0pt}\ {\isasympartial}kernel{\isacharunderscore}{\kern0pt}measure\ {\isacharparenleft}{\kern0pt}K\ {\isacharparenleft}{\kern0pt}Suc\ n{\isacharparenright}{\kern0pt}{\isacharparenright}{\kern0pt}\ {\isacharparenleft}{\kern0pt}{\isasymomega}\isactrlsub f\ n{\isacharparenright}{\kern0pt}{\isacharparenright}{\kern0pt}\isanewline
\ \ \ \ \ \ \ \ \ \ \ \ \ \ \ \ {\isasympartial}kernel{\isacharunderscore}{\kern0pt}measure\ {\isacharparenleft}{\kern0pt}PiK\ n\ K{\isacharparenright}{\kern0pt}\ {\isasymomega}{\isacharparenright}{\kern0pt}{\isachardoublequoteclose}%
}%
\DefineSnippet{kernel_product_convolution}{
\isacommand{theorem}\isamarkupfalse%
\ {\isacharparenleft}{\kern0pt}\isakeyword{in}\ prob{\isacharunderscore}{\kern0pt}space{\isacharparenright}{\kern0pt}\ kernel{\isacharunderscore}{\kern0pt}product{\isacharunderscore}{\kern0pt}convolution{\isacharcolon}{\kern0pt}\isanewline
\ \ \isakeyword{fixes}\ X\ {\isacharcolon}{\kern0pt}{\isacharcolon}{\kern0pt}\ {\isachardoublequoteopen}nat\ {\isasymRightarrow}\ {\isacharprime}{\kern0pt}a\ {\isasymRightarrow}\ {\isacharparenleft}{\kern0pt}{\isacharprime}{\kern0pt}b\ {\isacharcolon}{\kern0pt}{\isacharcolon}{\kern0pt}\ ordered{\isacharunderscore}{\kern0pt}euclidean{\isacharunderscore}{\kern0pt}space{\isacharparenright}{\kern0pt}{\isachardoublequoteclose}\isanewline
\ \ \isakeyword{assumes}\ {\isachardoublequoteopen}indep{\isacharunderscore}{\kern0pt}vars\ {\isacharparenleft}{\kern0pt}{\isasymlambda}{\isacharunderscore}{\kern0pt}{\isachardot}{\kern0pt}\ borel{\isacharparenright}{\kern0pt}\ X\ {\isacharbraceleft}{\kern0pt}{\isadigit{0}}{\isachardot}{\kern0pt}{\isachardot}{\kern0pt}{\isacharparenleft}{\kern0pt}n{\isacharcolon}{\kern0pt}{\isacharcolon}{\kern0pt}nat{\isacharparenright}{\kern0pt}{\isacharbraceright}{\kern0pt}{\isachardoublequoteclose}\isanewline
\ \ \isakeyword{defines}\ {\isachardoublequoteopen}S\ {\isasymequiv}\ {\isacharparenleft}{\kern0pt}{\isasymlambda}k\ {\isasymomega}{\isachardot}{\kern0pt}\ {\isasymSum}j\ {\isasymin}\ {\isacharbraceleft}{\kern0pt}{\isadigit{0}}{\isachardot}{\kern0pt}{\isachardot}{\kern0pt}k{\isacharbraceright}{\kern0pt}{\isachardot}{\kern0pt}\ X\ j\ {\isasymomega}{\isacharparenright}{\kern0pt}{\isachardoublequoteclose}\isanewline
\ \ \ \ \isakeyword{and}\ {\isachardoublequoteopen}K\ {\isasymequiv}\ {\isacharparenleft}{\kern0pt}{\isasymlambda}k{\isachardot}{\kern0pt}\ kernel{\isacharunderscore}{\kern0pt}of\ borel\ borel\ {\isacharparenleft}{\kern0pt}{\isasymlambda}x\ A{\isacharprime}{\kern0pt}{\isachardot}{\kern0pt}\ {\isacharparenleft}{\kern0pt}{\isacharparenleft}{\kern0pt}return\ borel\ x{\isacharparenright}{\kern0pt}\ {\isasymstar}\ {\isacharparenleft}{\kern0pt}distr\ M\ borel\ {\isacharparenleft}{\kern0pt}X\ k{\isacharparenright}{\kern0pt}{\isacharparenright}{\kern0pt}{\isacharparenright}{\kern0pt}\ A{\isacharprime}{\kern0pt}{\isacharparenright}{\kern0pt}{\isacharparenright}{\kern0pt}{\isachardoublequoteclose}\isanewline
\ \ \isakeyword{shows}\ {\isachardoublequoteopen}kernel{\isacharunderscore}{\kern0pt}measure\ {\isacharparenleft}{\kern0pt}PiK\ n\ K{\isacharparenright}{\kern0pt}\ {\isadigit{0}}\ {\isacharequal}{\kern0pt}\ {\isacharparenleft}{\kern0pt}{\isasymPi}\isactrlsub M\ t{\isasymin}{\isacharbraceleft}{\kern0pt}{\isadigit{0}}{\isachardot}{\kern0pt}{\isachardot}{\kern0pt}n{\isacharbraceright}{\kern0pt}{\isachardot}{\kern0pt}\ distr\ M\ borel\ {\isacharparenleft}{\kern0pt}X\ t{\isacharparenright}{\kern0pt}{\isacharparenright}{\kern0pt}{\isachardoublequoteclose}%
}%
\DefineSnippet{floor_pow2_lim}{
\isacommand{lemma}\isamarkupfalse%
\ floor{\isacharunderscore}{\kern0pt}pow{\isadigit{2}}{\isacharunderscore}{\kern0pt}lim{\isacharcolon}{\kern0pt}\ {\isachardoublequoteopen}{\isacharparenleft}{\kern0pt}{\isasymlambda}n{\isachardot}{\kern0pt}\ {\isasymlfloor}{\isadigit{2}}{\isacharcircum}{\kern0pt}n\ {\isacharasterisk}{\kern0pt}\ T{\isasymrfloor}\ {\isacharslash}{\kern0pt}\ {\isadigit{2}}\ {\isacharcircum}{\kern0pt}\ n{\isacharparenright}{\kern0pt}\ {\isasymlonglonglongrightarrow}\ T{\isachardoublequoteclose}%
}%
\DefineSnippet{dyadic_interval_step}{
\isacommand{definition}\isamarkupfalse%
\ dyadic{\isacharunderscore}{\kern0pt}interval{\isacharunderscore}{\kern0pt}step\ {\isacharcolon}{\kern0pt}{\isacharcolon}{\kern0pt}\ {\isachardoublequoteopen}nat\ {\isasymRightarrow}\ real\ {\isasymRightarrow}\ real\ {\isasymRightarrow}\ real\ set{\isachardoublequoteclose}\isanewline
\ \ \isakeyword{where}\ {\isachardoublequoteopen}dyadic{\isacharunderscore}{\kern0pt}interval{\isacharunderscore}{\kern0pt}step\ n\ S\ T\ {\isasymequiv}\ {\isacharparenleft}{\kern0pt}{\isasymlambda}k{\isachardot}{\kern0pt}\ k\ {\isacharslash}{\kern0pt}\ {\isacharparenleft}{\kern0pt}{\isadigit{2}}\ {\isacharcircum}{\kern0pt}\ n{\isacharparenright}{\kern0pt}{\isacharparenright}{\kern0pt}\ {\isacharbackquote}{\kern0pt}\ {\isacharbraceleft}{\kern0pt}{\isasymlceil}{\isadigit{2}}{\isacharcircum}{\kern0pt}n\ {\isacharasterisk}{\kern0pt}\ S{\isasymrceil}{\isachardot}{\kern0pt}{\isachardot}{\kern0pt}{\isasymlfloor}{\isadigit{2}}{\isacharcircum}{\kern0pt}n\ {\isacharasterisk}{\kern0pt}\ T{\isasymrfloor}{\isacharbraceright}{\kern0pt}{\isachardoublequoteclose}%
}%
\DefineSnippet{dyadic_interval}{
\isacommand{definition}\isamarkupfalse%
\ dyadic{\isacharunderscore}{\kern0pt}interval\ {\isacharcolon}{\kern0pt}{\isacharcolon}{\kern0pt}\ {\isachardoublequoteopen}real\ {\isasymRightarrow}\ real\ {\isasymRightarrow}\ real\ set{\isachardoublequoteclose}\isanewline
\ \ \isakeyword{where}\ {\isachardoublequoteopen}dyadic{\isacharunderscore}{\kern0pt}interval\ S\ T\ {\isasymequiv}\ {\isacharparenleft}{\kern0pt}{\isasymUnion}n{\isachardot}{\kern0pt}\ dyadic{\isacharunderscore}{\kern0pt}interval{\isacharunderscore}{\kern0pt}step\ n\ S\ T{\isacharparenright}{\kern0pt}{\isachardoublequoteclose}%
}%
\DefineSnippet{dyadic_interval_dense}{
\isacommand{lemma}\isamarkupfalse%
\ dyadic{\isacharunderscore}{\kern0pt}interval{\isacharunderscore}{\kern0pt}dense{\isacharcolon}{\kern0pt}\ {\isachardoublequoteopen}closure\ {\isacharparenleft}{\kern0pt}dyadic{\isacharunderscore}{\kern0pt}interval\ {\isadigit{0}}\ T{\isacharparenright}{\kern0pt}\ {\isacharequal}{\kern0pt}\ {\isacharbraceleft}{\kern0pt}{\isadigit{0}}{\isachardot}{\kern0pt}{\isachardot}{\kern0pt}T{\isacharbraceright}{\kern0pt}{\isachardoublequoteclose}%
}%
\DefineSnippet{dyadic_interval_countable}{
\isacommand{lemma}\isamarkupfalse%
\ dyadic{\isacharunderscore}{\kern0pt}interval{\isacharunderscore}{\kern0pt}countable{\isacharbrackleft}{\kern0pt}simp{\isacharbrackright}{\kern0pt}{\isacharcolon}{\kern0pt}\ {\isachardoublequoteopen}countable\ {\isacharparenleft}{\kern0pt}dyadic{\isacharunderscore}{\kern0pt}interval\ S\ T{\isacharparenright}{\kern0pt}{\isachardoublequoteclose}%
}%
\DefineSnippet{dyadic_interval_step_subset}{
\isacommand{lemma}\isamarkupfalse%
\ dyadic{\isacharunderscore}{\kern0pt}interval{\isacharunderscore}{\kern0pt}step{\isacharunderscore}{\kern0pt}subset{\isacharcolon}{\kern0pt}\isanewline
\ \ {\isachardoublequoteopen}n\ {\isasymle}\ m\ {\isasymLongrightarrow}\ dyadic{\isacharunderscore}{\kern0pt}interval{\isacharunderscore}{\kern0pt}step\ n\ {\isadigit{0}}\ T\ {\isasymsubseteq}\ dyadic{\isacharunderscore}{\kern0pt}interval{\isacharunderscore}{\kern0pt}step\ m\ {\isadigit{0}}\ T{\isachardoublequoteclose}%
}%
\DefineSnippet{dyadic_expansion}{
\isacommand{definition}\isamarkupfalse%
\ {\isachardoublequoteopen}dyadic{\isacharunderscore}{\kern0pt}expansion\ x\ n\ b\ k\ {\isasymequiv}\ set\ b\ {\isasymsubseteq}\ {\isacharbraceleft}{\kern0pt}{\isadigit{0}}{\isacharcomma}{\kern0pt}{\isadigit{1}}{\isacharbraceright}{\kern0pt}\isanewline
\ \ {\isasymand}\ length\ b\ {\isacharequal}{\kern0pt}\ n\ {\isasymand}\ x\ {\isacharequal}{\kern0pt}\ real{\isacharunderscore}{\kern0pt}of{\isacharunderscore}{\kern0pt}int\ k\ {\isacharplus}{\kern0pt}\ {\isacharparenleft}{\kern0pt}{\isasymSum}m{\isasymin}{\isacharbraceleft}{\kern0pt}{\isadigit{1}}{\isachardot}{\kern0pt}{\isachardot}{\kern0pt}n{\isacharbraceright}{\kern0pt}{\isachardot}{\kern0pt}\ real\ {\isacharparenleft}{\kern0pt}b\ {\isacharbang}{\kern0pt}\ {\isacharparenleft}{\kern0pt}m{\isacharminus}{\kern0pt}{\isadigit{1}}{\isacharparenright}{\kern0pt}{\isacharparenright}{\kern0pt}\ {\isacharslash}{\kern0pt}\ {\isadigit{2}}\ {\isacharcircum}{\kern0pt}\ m{\isacharparenright}{\kern0pt}{\isachardoublequoteclose}%
}%
\DefineSnippet{dyadic_expansion_exists}{
\isacommand{lemma}\isamarkupfalse%
\ dyadic{\isacharunderscore}{\kern0pt}expansion{\isacharunderscore}{\kern0pt}ex{\isacharcolon}{\kern0pt}\isanewline
\ \ \isakeyword{assumes}\ {\isachardoublequoteopen}x\ {\isasymin}\ dyadic{\isacharunderscore}{\kern0pt}interval{\isacharunderscore}{\kern0pt}step\ n\ {\isadigit{0}}\ T{\isachardoublequoteclose}\ \isanewline
\ \ \isakeyword{shows}\ {\isachardoublequoteopen}{\isasymexists}b\ k{\isachardot}{\kern0pt}\ dyadic{\isacharunderscore}{\kern0pt}expansion\ x\ n\ b\ k{\isachardoublequoteclose}%
}%
\DefineSnippet{stochastic_process}{
\isacommand{locale}\isamarkupfalse%
\ stochastic{\isacharunderscore}{\kern0pt}process\ {\isacharequal}{\kern0pt}\ prob{\isacharunderscore}{\kern0pt}space\ {\isacharplus}{\kern0pt}\isanewline
\ \ \isakeyword{fixes}\ M{\isacharprime}{\kern0pt}\ {\isacharcolon}{\kern0pt}{\isacharcolon}{\kern0pt}\ {\isachardoublequoteopen}{\isacharprime}{\kern0pt}b\ measure{\isachardoublequoteclose}\isanewline
\ \ \ \ \isakeyword{and}\ I\ {\isacharcolon}{\kern0pt}{\isacharcolon}{\kern0pt}\ {\isachardoublequoteopen}{\isacharprime}{\kern0pt}t\ set{\isachardoublequoteclose}\isanewline
\ \ \ \ \isakeyword{and}\ X\ {\isacharcolon}{\kern0pt}{\isacharcolon}{\kern0pt}\ {\isachardoublequoteopen}{\isacharprime}{\kern0pt}t\ {\isasymRightarrow}\ {\isacharprime}{\kern0pt}a\ {\isasymRightarrow}\ {\isacharprime}{\kern0pt}b{\isachardoublequoteclose}\isanewline
\ \ \isakeyword{assumes}\ random{\isacharunderscore}{\kern0pt}process{\isacharbrackleft}{\kern0pt}measurable{\isacharbrackright}{\kern0pt}{\isacharcolon}{\kern0pt}\ {\isachardoublequoteopen}{\isasymAnd}i{\isachardot}{\kern0pt}\ random{\isacharunderscore}{\kern0pt}variable\ M{\isacharprime}{\kern0pt}\ {\isacharparenleft}{\kern0pt}X\ i{\isacharparenright}{\kern0pt}{\isachardoublequoteclose}%
}%
\DefineSnippet{stochastic_process_typedef}{
\isacommand{typedef}\isamarkupfalse%
\ {\isacharparenleft}{\kern0pt}{\isacharprime}{\kern0pt}t{\isacharcomma}{\kern0pt}\ {\isacharprime}{\kern0pt}a{\isacharcomma}{\kern0pt}\ {\isacharprime}{\kern0pt}b{\isacharparenright}{\kern0pt}\ stochastic{\isacharunderscore}{\kern0pt}process\ {\isacharequal}{\kern0pt}\isanewline
\ {\isachardoublequoteopen}{\isacharbraceleft}{\kern0pt}{\isacharparenleft}{\kern0pt}M\ {\isacharcolon}{\kern0pt}{\isacharcolon}{\kern0pt}\ {\isacharprime}{\kern0pt}a\ measure{\isacharcomma}{\kern0pt}\ M{\isacharprime}{\kern0pt}\ {\isacharcolon}{\kern0pt}{\isacharcolon}{\kern0pt}\ {\isacharprime}{\kern0pt}b\ measure{\isacharcomma}{\kern0pt}\ I\ {\isacharcolon}{\kern0pt}{\isacharcolon}{\kern0pt}\ {\isacharprime}{\kern0pt}t\ set{\isacharcomma}{\kern0pt}\ X\ {\isacharcolon}{\kern0pt}{\isacharcolon}{\kern0pt}\ {\isacharprime}{\kern0pt}t\ {\isasymRightarrow}\ {\isacharprime}{\kern0pt}a\ {\isasymRightarrow}\ {\isacharprime}{\kern0pt}b{\isacharparenright}{\kern0pt}{\isachardot}{\kern0pt}\isanewline
\ \ \ stochastic{\isacharunderscore}{\kern0pt}process\ M\ M{\isacharprime}{\kern0pt}\ X{\isacharbraceright}{\kern0pt}{\isachardoublequoteclose}%
}%
\DefineSnippet{distributions}{
\isacommand{lift{\isacharunderscore}{\kern0pt}definition}\isamarkupfalse%
\ distributions\ {\isacharcolon}{\kern0pt}{\isacharcolon}{\kern0pt}\ {\isachardoublequoteopen}{\isacharparenleft}{\kern0pt}{\isacharprime}{\kern0pt}t{\isacharcomma}{\kern0pt}\ {\isacharprime}{\kern0pt}a{\isacharcomma}{\kern0pt}\ {\isacharprime}{\kern0pt}b{\isacharparenright}{\kern0pt}\ stochastic{\isacharunderscore}{\kern0pt}process\ {\isasymRightarrow}\ {\isacharprime}{\kern0pt}t\ set\ {\isasymRightarrow}\ {\isacharparenleft}{\kern0pt}{\isacharprime}{\kern0pt}t\ {\isasymRightarrow}\ {\isacharprime}{\kern0pt}b{\isacharparenright}{\kern0pt}\ measure{\isachardoublequoteclose}\isanewline
\isakeyword{is}\ {\isachardoublequoteopen}{\isasymlambda}{\isacharparenleft}{\kern0pt}M{\isacharcomma}{\kern0pt}\ M{\isacharprime}{\kern0pt}{\isacharcomma}{\kern0pt}\ {\isacharunderscore}{\kern0pt}{\isacharcomma}{\kern0pt}\ X{\isacharparenright}{\kern0pt}\ T{\isachardot}{\kern0pt}\ {\isacharparenleft}{\kern0pt}{\isasymPi}\isactrlsub M\ t{\isasymin}T{\isachardot}{\kern0pt}\ distr\ M\ M{\isacharprime}{\kern0pt}\ {\isacharparenleft}{\kern0pt}X\ t{\isacharparenright}{\kern0pt}{\isacharparenright}{\kern0pt}{\isachardoublequoteclose}%
}%
\DefineSnippet{indep_increments}{
\isacommand{lift{\isacharunderscore}{\kern0pt}definition}\isamarkupfalse%
\ indep{\isacharunderscore}{\kern0pt}increments\ {\isacharcolon}{\kern0pt}{\isacharcolon}{\kern0pt}\ {\isachardoublequoteopen}{\isacharparenleft}{\kern0pt}{\isacharprime}{\kern0pt}t\ {\isacharcolon}{\kern0pt}{\isacharcolon}{\kern0pt}\ linorder{\isacharcomma}{\kern0pt}\ {\isacharprime}{\kern0pt}a{\isacharcomma}{\kern0pt}\ {\isacharprime}{\kern0pt}b\ {\isacharcolon}{\kern0pt}{\isacharcolon}{\kern0pt}\ minus{\isacharparenright}{\kern0pt}\ stochastic{\isacharunderscore}{\kern0pt}process\ {\isasymRightarrow}\ bool{\isachardoublequoteclose}\ \isakeyword{is}\isanewline
{\isachardoublequoteopen}{\isasymlambda}{\isacharparenleft}{\kern0pt}M{\isacharcomma}{\kern0pt}\ M{\isacharprime}{\kern0pt}{\isacharcomma}{\kern0pt}\ I{\isacharcomma}{\kern0pt}\ X{\isacharparenright}{\kern0pt}{\isachardot}{\kern0pt}\isanewline
\ \ {\isacharparenleft}{\kern0pt}{\isasymforall}l{\isachardot}{\kern0pt}\ set\ l\ {\isasymsubseteq}\ I\ {\isasymand}\ sorted\ l\ {\isasymand}\ length\ l\ {\isasymge}\ {\isadigit{2}}\ {\isasymlongrightarrow}\isanewline
\ \ \ \ \ prob{\isacharunderscore}{\kern0pt}space{\isachardot}{\kern0pt}indep{\isacharunderscore}{\kern0pt}vars\ M\ {\isacharparenleft}{\kern0pt}{\isasymlambda}{\isacharunderscore}{\kern0pt}{\isachardot}{\kern0pt}\ M{\isacharprime}{\kern0pt}{\isacharparenright}{\kern0pt}\ {\isacharparenleft}{\kern0pt}{\isasymlambda}k\ v{\isachardot}{\kern0pt}\ X\ {\isacharparenleft}{\kern0pt}l{\isacharbang}{\kern0pt}k{\isacharparenright}{\kern0pt}\ v\ {\isacharminus}{\kern0pt}\ X\ {\isacharparenleft}{\kern0pt}l{\isacharbang}{\kern0pt}{\isacharparenleft}{\kern0pt}k{\isacharminus}{\kern0pt}{\isadigit{1}}{\isacharparenright}{\kern0pt}{\isacharparenright}{\kern0pt}\ v{\isacharparenright}{\kern0pt}\ {\isacharbraceleft}{\kern0pt}{\isadigit{1}}{\isachardot}{\kern0pt}{\isachardot}{\kern0pt}{\isacharless}{\kern0pt}length\ l{\isacharbraceright}{\kern0pt}{\isacharparenright}{\kern0pt}{\isachardoublequoteclose}%
\isadelimproof
\ %
\endisadelimproof
\isatagproof
\isacommand{{\isachardot}{\kern0pt}}\isamarkupfalse%
\endisatagproof
{\isafoldproof}%
\isadelimproof
\endisadelimproof
}%
\DefineSnippet{stationary_increments}{
\isacommand{definition}\isamarkupfalse%
\ {\isachardoublequoteopen}stationary{\isacharunderscore}{\kern0pt}increments\ X\ {\isasymlongleftrightarrow}\ {\isacharparenleft}{\kern0pt}{\isasymforall}t{\isadigit{1}}\ t{\isadigit{2}}\ k{\isachardot}{\kern0pt}\ t{\isadigit{1}}\ {\isachargreater}{\kern0pt}\ {\isadigit{0}}\ {\isasymand}\ t{\isadigit{2}}\ {\isachargreater}{\kern0pt}\ {\isadigit{0}}\ {\isasymand}\ k\ {\isachargreater}{\kern0pt}\ {\isadigit{0}}\ {\isasymlongrightarrow}\ \isanewline
\ \ \ \ \ distr\ {\isacharparenleft}{\kern0pt}proc{\isacharunderscore}{\kern0pt}source\ X{\isacharparenright}{\kern0pt}\ {\isacharparenleft}{\kern0pt}proc{\isacharunderscore}{\kern0pt}target\ X{\isacharparenright}{\kern0pt}\ {\isacharparenleft}{\kern0pt}{\isasymlambda}v{\isachardot}{\kern0pt}\ X\ {\isacharparenleft}{\kern0pt}t{\isadigit{1}}\ {\isacharplus}{\kern0pt}\ k{\isacharparenright}{\kern0pt}\ v\ {\isacharminus}{\kern0pt}\ X\ t{\isadigit{1}}\ v{\isacharparenright}{\kern0pt}\ {\isacharequal}{\kern0pt}\isanewline
\ \ \ \ \ distr\ {\isacharparenleft}{\kern0pt}proc{\isacharunderscore}{\kern0pt}source\ X{\isacharparenright}{\kern0pt}\ {\isacharparenleft}{\kern0pt}proc{\isacharunderscore}{\kern0pt}target\ X{\isacharparenright}{\kern0pt}\ {\isacharparenleft}{\kern0pt}{\isasymlambda}v{\isachardot}{\kern0pt}\ X\ {\isacharparenleft}{\kern0pt}t{\isadigit{2}}\ {\isacharplus}{\kern0pt}\ k{\isacharparenright}{\kern0pt}\ v\ {\isacharminus}{\kern0pt}\ X\ t{\isadigit{2}}\ v{\isacharparenright}{\kern0pt}{\isacharparenright}{\kern0pt}{\isachardoublequoteclose}%
}%
\DefineSnippet{compatible}{
\isacommand{lift{\isacharunderscore}{\kern0pt}definition}\isamarkupfalse%
\ compatible\ {\isacharcolon}{\kern0pt}{\isacharcolon}{\kern0pt}\ {\isachardoublequoteopen}{\isacharparenleft}{\kern0pt}{\isacharprime}{\kern0pt}t{\isacharcomma}{\kern0pt}{\isacharprime}{\kern0pt}a{\isacharcomma}{\kern0pt}{\isacharprime}{\kern0pt}b{\isacharparenright}{\kern0pt}\ stochastic{\isacharunderscore}{\kern0pt}process\ {\isasymRightarrow}\ {\isacharparenleft}{\kern0pt}{\isacharprime}{\kern0pt}t{\isacharcomma}{\kern0pt}{\isacharprime}{\kern0pt}a{\isacharcomma}{\kern0pt}{\isacharprime}{\kern0pt}b{\isacharparenright}{\kern0pt}\ stochastic{\isacharunderscore}{\kern0pt}process\ {\isasymRightarrow}\ bool{\isachardoublequoteclose}\isanewline
\isakeyword{is}\ {\isachardoublequoteopen}{\isasymlambda}{\isacharparenleft}{\kern0pt}Mx{\isacharcomma}{\kern0pt}\ M{\isacharprime}{\kern0pt}x{\isacharcomma}{\kern0pt}\ Ix{\isacharcomma}{\kern0pt}\ X{\isacharparenright}{\kern0pt}\ {\isacharparenleft}{\kern0pt}My{\isacharcomma}{\kern0pt}\ M{\isacharprime}{\kern0pt}y{\isacharcomma}{\kern0pt}\ Iy{\isacharcomma}{\kern0pt}\ {\isacharunderscore}{\kern0pt}{\isacharparenright}{\kern0pt}{\isachardot}{\kern0pt}\ Mx\ {\isacharequal}{\kern0pt}\ My\ {\isasymand}\ sets\ M{\isacharprime}{\kern0pt}x\ {\isacharequal}{\kern0pt}\ sets\ M{\isacharprime}{\kern0pt}y\ {\isasymand}\ Ix\ {\isacharequal}{\kern0pt}\ Iy{\isachardoublequoteclose}%
\isadelimproof
\ %
\endisadelimproof
\isatagproof
\isacommand{{\isachardot}{\kern0pt}}\isamarkupfalse%
\endisatagproof
{\isafoldproof}%
\isadelimproof
\endisadelimproof
}%
\DefineSnippet{modification}{
\isacommand{definition}\isamarkupfalse%
\ modification\ {\isacharcolon}{\kern0pt}{\isacharcolon}{\kern0pt}\ {\isachardoublequoteopen}{\isacharparenleft}{\kern0pt}{\isacharprime}{\kern0pt}t{\isacharcomma}{\kern0pt}{\isacharprime}{\kern0pt}a{\isacharcomma}{\kern0pt}{\isacharprime}{\kern0pt}b{\isacharparenright}{\kern0pt}\ stochastic{\isacharunderscore}{\kern0pt}process\ {\isasymRightarrow}\ {\isacharparenleft}{\kern0pt}{\isacharprime}{\kern0pt}t{\isacharcomma}{\kern0pt}{\isacharprime}{\kern0pt}a{\isacharcomma}{\kern0pt}{\isacharprime}{\kern0pt}b{\isacharparenright}{\kern0pt}\ stochastic{\isacharunderscore}{\kern0pt}process\ {\isasymRightarrow}\ bool{\isachardoublequoteclose}\ \isakeyword{where}\isanewline
{\isachardoublequoteopen}modification\ X\ Y\ {\isasymlongleftrightarrow}\ compatible\ X\ Y\ {\isasymand}\ {\isacharparenleft}{\kern0pt}{\isasymforall}t\ {\isasymin}\ proc{\isacharunderscore}{\kern0pt}index\ X{\isachardot}{\kern0pt}\ AE\ x\ in\ proc{\isacharunderscore}{\kern0pt}source\ X{\isachardot}{\kern0pt}\ X\ t\ x\ {\isacharequal}{\kern0pt}\ Y\ t\ x{\isacharparenright}{\kern0pt}{\isachardoublequoteclose}%
}%
\DefineSnippet{modification_sym}{
\isacommand{lemma}\isamarkupfalse%
\ modification{\isacharunderscore}{\kern0pt}sym{\isacharcolon}{\kern0pt}\ {\isachardoublequoteopen}modification\ X\ Y\ {\isasymLongrightarrow}\ modification\ Y\ X{\isachardoublequoteclose}\isanewline
\isadelimproof
\endisadelimproof
\isatagproof
\isacommand{proof}\isamarkupfalse%
\ {\isacharparenleft}{\kern0pt}rule\ modificationI{\isacharcomma}{\kern0pt}\ safe{\isacharparenright}{\kern0pt}\isanewline
\ \ \isacommand{assume}\isamarkupfalse%
\ {\isacharasterisk}{\kern0pt}{\isacharcolon}{\kern0pt}\ {\isachardoublequoteopen}modification\ X\ Y{\isachardoublequoteclose}\isanewline
\ \ \isacommand{then}\isamarkupfalse%
\ \isacommand{show}\isamarkupfalse%
\ compat{\isacharcolon}{\kern0pt}\ {\isachardoublequoteopen}compatible\ Y\ X{\isachardoublequoteclose}\isanewline
\ \ \ \ \isacommand{using}\isamarkupfalse%
\ compatible{\isacharunderscore}{\kern0pt}sym\ modificationD{\isacharparenleft}{\kern0pt}{\isadigit{1}}{\isacharparenright}{\kern0pt}\ \isacommand{by}\isamarkupfalse%
\ blast\isanewline
\ \ \isacommand{fix}\isamarkupfalse%
\ t\ \isacommand{assume}\isamarkupfalse%
\ {\isachardoublequoteopen}t\ {\isasymin}\ proc{\isacharunderscore}{\kern0pt}index\ Y{\isachardoublequoteclose}\isanewline
\ \ \isacommand{then}\isamarkupfalse%
\ \isacommand{have}\isamarkupfalse%
\ {\isachardoublequoteopen}t\ {\isasymin}\ proc{\isacharunderscore}{\kern0pt}index\ X{\isachardoublequoteclose}\isanewline
\ \ \ \ \isacommand{using}\isamarkupfalse%
\ compatible{\isacharunderscore}{\kern0pt}index{\isacharbrackleft}{\kern0pt}OF\ compat{\isacharbrackright}{\kern0pt}\ \isacommand{by}\isamarkupfalse%
\ blast\isanewline
\ \ \isacommand{have}\isamarkupfalse%
\ {\isachardoublequoteopen}AE\ x\ in\ proc{\isacharunderscore}{\kern0pt}source\ Y{\isachardot}{\kern0pt}\ X\ t\ x\ {\isacharequal}{\kern0pt}\ Y\ t\ x{\isachardoublequoteclose}\isanewline
\ \ \ \ \isacommand{using}\isamarkupfalse%
\ modificationD{\isacharparenleft}{\kern0pt}{\isadigit{2}}{\isacharparenright}{\kern0pt}{\isacharbrackleft}{\kern0pt}OF\ {\isacharasterisk}{\kern0pt}\ {\isacartoucheopen}t\ {\isasymin}\ proc{\isacharunderscore}{\kern0pt}index\ X{\isacartoucheclose}{\isacharbrackright}{\kern0pt}\isanewline
\ \ \ \ \ \ \ \ \ \ compatible{\isacharunderscore}{\kern0pt}source{\isacharbrackleft}{\kern0pt}OF\ compat{\isacharbrackright}{\kern0pt}\ \isacommand{by}\isamarkupfalse%
\ argo\isanewline
\ \ \isacommand{then}\isamarkupfalse%
\ \isacommand{show}\isamarkupfalse%
\ {\isachardoublequoteopen}AE\ x\ in\ proc{\isacharunderscore}{\kern0pt}source\ Y{\isachardot}{\kern0pt}\ Y\ t\ x\ {\isacharequal}{\kern0pt}\ X\ t\ x{\isachardoublequoteclose}\isanewline
\ \ \ \ \isacommand{by}\isamarkupfalse%
\ force\isanewline
\isacommand{qed}\isamarkupfalse%
\endisatagproof
{\isafoldproof}%
\isadelimproof
\endisadelimproof
}%
\DefineSnippet{modification_imp_identical_distributions}{
\isacommand{lemma}\isamarkupfalse%
\ modification{\isacharunderscore}{\kern0pt}imp{\isacharunderscore}{\kern0pt}identical{\isacharunderscore}{\kern0pt}distributions{\isacharcolon}{\kern0pt}\isanewline
\ \ \isakeyword{assumes}\ modification{\isacharcolon}{\kern0pt}\ {\isachardoublequoteopen}modification\ X\ Y{\isachardoublequoteclose}\isanewline
\ \ \ \ \ \ \isakeyword{and}\ index{\isacharcolon}{\kern0pt}\ {\isachardoublequoteopen}T\ {\isasymsubseteq}\ proc{\isacharunderscore}{\kern0pt}index\ X{\isachardoublequoteclose}\isanewline
\ \ \isakeyword{shows}\ {\isachardoublequoteopen}distributions\ X\ T\ {\isacharequal}{\kern0pt}\ distributions\ Y\ T{\isachardoublequoteclose}\isanewline
\isadelimproof
\endisadelimproof
\isatagproof
\isacommand{proof}\isamarkupfalse%
\ {\isacharminus}{\kern0pt}\isanewline
\ \ \isacommand{have}\isamarkupfalse%
\ {\isachardoublequoteopen}proc{\isacharunderscore}{\kern0pt}source\ X\ {\isacharequal}{\kern0pt}\ proc{\isacharunderscore}{\kern0pt}source\ Y{\isachardoublequoteclose}\isanewline
\ \ \ \ \isacommand{by}\isamarkupfalse%
\ {\isacharparenleft}{\kern0pt}simp\ add{\isacharcolon}{\kern0pt}\ modification\ modificationD{\isacharparenleft}{\kern0pt}{\isadigit{1}}{\isacharparenright}{\kern0pt}{\isacharparenright}{\kern0pt}\isanewline
\ \ \isacommand{moreover}\isamarkupfalse%
\ \isacommand{have}\isamarkupfalse%
\ {\isachardoublequoteopen}sets\ {\isacharparenleft}{\kern0pt}proc{\isacharunderscore}{\kern0pt}target\ X{\isacharparenright}{\kern0pt}\ {\isacharequal}{\kern0pt}\ sets\ {\isacharparenleft}{\kern0pt}proc{\isacharunderscore}{\kern0pt}target\ Y{\isacharparenright}{\kern0pt}{\isachardoublequoteclose}\isanewline
\ \ \ \ \isacommand{by}\isamarkupfalse%
\ {\isacharparenleft}{\kern0pt}simp\ add{\isacharcolon}{\kern0pt}\ modification\ modificationD{\isacharparenleft}{\kern0pt}{\isadigit{1}}{\isacharparenright}{\kern0pt}{\isacharparenright}{\kern0pt}\isanewline
\ \ \isacommand{ultimately}\isamarkupfalse%
\ \isacommand{have}\isamarkupfalse%
\ {\isachardoublequoteopen}distr\ {\isacharparenleft}{\kern0pt}proc{\isacharunderscore}{\kern0pt}source\ X{\isacharparenright}{\kern0pt}\ {\isacharparenleft}{\kern0pt}proc{\isacharunderscore}{\kern0pt}target\ X{\isacharparenright}{\kern0pt}\ {\isacharparenleft}{\kern0pt}X\ x{\isacharparenright}{\kern0pt}\ {\isacharequal}{\kern0pt}\isanewline
\ \ \ \ \ \ \ \ \ distr\ {\isacharparenleft}{\kern0pt}proc{\isacharunderscore}{\kern0pt}source\ Y{\isacharparenright}{\kern0pt}\ {\isacharparenleft}{\kern0pt}proc{\isacharunderscore}{\kern0pt}target\ Y{\isacharparenright}{\kern0pt}\ {\isacharparenleft}{\kern0pt}Y\ x{\isacharparenright}{\kern0pt}{\isachardoublequoteclose}\isanewline
\ \ \ \ \ \ \ \isakeyword{if}\ {\isachardoublequoteopen}x\ {\isasymin}\ T{\isachardoublequoteclose}\ \isakeyword{for}\ x\isanewline
\ \ \ \ \isacommand{apply}\isamarkupfalse%
\ {\isacharparenleft}{\kern0pt}rule\ distr{\isacharunderscore}{\kern0pt}cong{\isacharunderscore}{\kern0pt}AE{\isacharparenright}{\kern0pt}\isanewline
\ \ \ \ \ \ \isacommand{apply}\isamarkupfalse%
\ {\isacharparenleft}{\kern0pt}metis\ assms\ modificationD{\isacharparenleft}{\kern0pt}{\isadigit{2}}{\isacharparenright}{\kern0pt}\ subset{\isacharunderscore}{\kern0pt}eq\ that{\isacharparenright}{\kern0pt}\isanewline
\ \ \ \ \ \isacommand{apply}\isamarkupfalse%
\ simp{\isacharunderscore}{\kern0pt}all\isanewline
\ \ \ \ \isacommand{done}\isamarkupfalse%
\isanewline
\ \ \isacommand{then}\isamarkupfalse%
\ \isacommand{show}\isamarkupfalse%
\ {\isacharquery}{\kern0pt}thesis\isanewline
\ \ \ \ \isacommand{by}\isamarkupfalse%
\ {\isacharparenleft}{\kern0pt}auto\ simp{\isacharcolon}{\kern0pt}\ distributions{\isacharunderscore}{\kern0pt}altdef\ cong{\isacharcolon}{\kern0pt}\ PiM{\isacharunderscore}{\kern0pt}cong{\isacharparenright}{\kern0pt}\isanewline
\isacommand{qed}\isamarkupfalse%
\endisatagproof
{\isafoldproof}%
\isadelimproof
\endisadelimproof
}%
\DefineSnippet{indistinguishable}{
\isacommand{definition}\isamarkupfalse%
\ indistinguishable\ {\isacharcolon}{\kern0pt}{\isacharcolon}{\kern0pt}\ {\isachardoublequoteopen}{\isacharparenleft}{\kern0pt}{\isacharprime}{\kern0pt}t{\isacharcomma}{\kern0pt}{\isacharprime}{\kern0pt}a{\isacharcomma}{\kern0pt}{\isacharprime}{\kern0pt}b{\isacharparenright}{\kern0pt}\ stochastic{\isacharunderscore}{\kern0pt}process\ {\isasymRightarrow}\ {\isacharparenleft}{\kern0pt}{\isacharprime}{\kern0pt}t{\isacharcomma}{\kern0pt}{\isacharprime}{\kern0pt}a{\isacharcomma}{\kern0pt}{\isacharprime}{\kern0pt}b{\isacharparenright}{\kern0pt}\ stochastic{\isacharunderscore}{\kern0pt}process\ {\isasymRightarrow}\ bool{\isachardoublequoteclose}\ \isakeyword{where}\isanewline
{\isachardoublequoteopen}indistinguishable\ X\ Y\ {\isasymlongleftrightarrow}\ compatible\ X\ Y\ {\isasymand}\isanewline
\ {\isacharparenleft}{\kern0pt}{\isasymexists}N\ {\isasymin}\ null{\isacharunderscore}{\kern0pt}sets\ {\isacharparenleft}{\kern0pt}proc{\isacharunderscore}{\kern0pt}source\ X{\isacharparenright}{\kern0pt}{\isachardot}{\kern0pt}\ {\isasymforall}t\ {\isasymin}\ proc{\isacharunderscore}{\kern0pt}index\ X{\isachardot}{\kern0pt}\ {\isacharbraceleft}{\kern0pt}x\ {\isasymin}\ space\ {\isacharparenleft}{\kern0pt}proc{\isacharunderscore}{\kern0pt}source\ X{\isacharparenright}{\kern0pt}{\isachardot}{\kern0pt}\ X\ t\ x\ {\isasymnoteq}\ Y\ t\ x{\isacharbraceright}{\kern0pt}\ {\isasymsubseteq}\ N{\isacharparenright}{\kern0pt}{\isachardoublequoteclose}%
}%
\DefineSnippet{modification_countable}{
\isacommand{lemma}\isamarkupfalse%
\ modification{\isacharunderscore}{\kern0pt}countable{\isacharcolon}{\kern0pt}\isanewline
\ \ \isakeyword{assumes}\ {\isachardoublequoteopen}modification\ X\ Y{\isachardoublequoteclose}\ {\isachardoublequoteopen}countable\ {\isacharparenleft}{\kern0pt}proc{\isacharunderscore}{\kern0pt}index\ X{\isacharparenright}{\kern0pt}{\isachardoublequoteclose}\isanewline
\ \ \isakeyword{shows}\ {\isachardoublequoteopen}indistinguishable\ X\ Y{\isachardoublequoteclose}%
}%
\DefineSnippet{modification_continuous_indistinguishable}{
\isacommand{lemma}\isamarkupfalse%
\ modification{\isacharunderscore}{\kern0pt}continuous{\isacharunderscore}{\kern0pt}indistinguishable{\isacharcolon}{\kern0pt}\isanewline
\ \ \isakeyword{fixes}\ X\ {\isacharcolon}{\kern0pt}{\isacharcolon}{\kern0pt}\ {\isachardoublequoteopen}{\isacharparenleft}{\kern0pt}real{\isacharcomma}{\kern0pt}\ {\isacharprime}{\kern0pt}a{\isacharcomma}{\kern0pt}\ {\isacharprime}{\kern0pt}b\ {\isacharcolon}{\kern0pt}{\isacharcolon}{\kern0pt}\ metric{\isacharunderscore}{\kern0pt}space{\isacharparenright}{\kern0pt}\ stochastic{\isacharunderscore}{\kern0pt}process{\isachardoublequoteclose}\isanewline
\ \ \isakeyword{assumes}\ modification{\isacharcolon}{\kern0pt}\ {\isachardoublequoteopen}modification\ X\ Y{\isachardoublequoteclose}\isanewline
\ \ \ \ \isakeyword{and}\ interval{\isacharcolon}{\kern0pt}\ {\isachardoublequoteopen}{\isasymexists}T\ {\isachargreater}{\kern0pt}\ {\isadigit{0}}{\isachardot}{\kern0pt}\ proc{\isacharunderscore}{\kern0pt}index\ X\ {\isacharequal}{\kern0pt}\ {\isacharbraceleft}{\kern0pt}{\isadigit{0}}{\isachardot}{\kern0pt}{\isachardot}{\kern0pt}T{\isacharbraceright}{\kern0pt}{\isachardoublequoteclose}\isanewline
\ \ \ \ \ \ \isakeyword{and}\ rc{\isacharcolon}{\kern0pt}\ {\isachardoublequoteopen}AE\ {\isasymomega}\ in\ proc{\isacharunderscore}{\kern0pt}source\ X{\isachardot}{\kern0pt}\ continuous{\isacharunderscore}{\kern0pt}on\ {\isacharparenleft}{\kern0pt}proc{\isacharunderscore}{\kern0pt}index\ X{\isacharparenright}{\kern0pt}\ {\isacharparenleft}{\kern0pt}{\isasymlambda}t{\isachardot}{\kern0pt}\ X\ t\ {\isasymomega}{\isacharparenright}{\kern0pt}{\isachardoublequoteclose}\isanewline
\ \ \ \ \ \ \ \ \ \ \ \ \ \ \ \ \ \ {\isacharparenleft}{\kern0pt}\isakeyword{is}\ {\isachardoublequoteopen}AE\ {\isasymomega}\ in\ proc{\isacharunderscore}{\kern0pt}source\ X{\isachardot}{\kern0pt}\ {\isacharquery}{\kern0pt}cont{\isacharunderscore}{\kern0pt}X\ {\isasymomega}{\isachardoublequoteclose}{\isacharparenright}{\kern0pt}\isanewline
\ \ \ \ \ \ \ \ \ \ \ \ \ \ {\isachardoublequoteopen}AE\ {\isasymomega}\ in\ proc{\isacharunderscore}{\kern0pt}source\ Y{\isachardot}{\kern0pt}\ continuous{\isacharunderscore}{\kern0pt}on\ {\isacharparenleft}{\kern0pt}proc{\isacharunderscore}{\kern0pt}index\ Y{\isacharparenright}{\kern0pt}\ {\isacharparenleft}{\kern0pt}{\isasymlambda}t{\isachardot}{\kern0pt}\ Y\ t\ {\isasymomega}{\isacharparenright}{\kern0pt}{\isachardoublequoteclose}\isanewline
\ \ \ \ \ \ \ \ \ \ \ \ \ \ \ \ \ \ {\isacharparenleft}{\kern0pt}\isakeyword{is}\ {\isachardoublequoteopen}AE\ {\isasymomega}\ in\ proc{\isacharunderscore}{\kern0pt}source\ Y{\isachardot}{\kern0pt}\ {\isacharquery}{\kern0pt}cont{\isacharunderscore}{\kern0pt}Y\ {\isasymomega}{\isachardoublequoteclose}{\isacharparenright}{\kern0pt}\isanewline
\ \ \isakeyword{shows}\ {\isachardoublequoteopen}indistinguishable\ X\ Y{\isachardoublequoteclose}%
}%
\DefineSnippet{transition_kernel}{
\isacommand{locale}\isamarkupfalse%
\ transition{\isacharunderscore}{\kern0pt}kernel\ {\isacharequal}{\kern0pt}\isanewline
\ \ \isakeyword{fixes}\ M\ {\isacharcolon}{\kern0pt}{\isacharcolon}{\kern0pt}\ {\isachardoublequoteopen}{\isacharprime}{\kern0pt}a\ measure{\isachardoublequoteclose}\isanewline
\ \ \ \ \isakeyword{and}\ M{\isacharprime}{\kern0pt}\ {\isacharcolon}{\kern0pt}{\isacharcolon}{\kern0pt}\ {\isachardoublequoteopen}{\isacharprime}{\kern0pt}b\ measure{\isachardoublequoteclose}\isanewline
\ \ \ \ \isakeyword{and}\ {\isasymkappa}\ {\isacharcolon}{\kern0pt}{\isacharcolon}{\kern0pt}\ {\isachardoublequoteopen}{\isacharprime}{\kern0pt}a\ {\isasymRightarrow}\ {\isacharprime}{\kern0pt}b\ set\ {\isasymRightarrow}\ ennreal{\isachardoublequoteclose}\isanewline
\ \ \isakeyword{assumes}\isanewline
\ \ \ \ sets{\isacharunderscore}{\kern0pt}target{\isacharunderscore}{\kern0pt}measurable\ {\isacharbrackleft}{\kern0pt}measurable{\isacharbrackright}{\kern0pt}{\isacharcolon}{\kern0pt}\isanewline
\ \ \ \ \ \ {\isachardoublequoteopen}{\isasymAnd}A{\isacharprime}{\kern0pt}{\isachardot}{\kern0pt}\ A{\isacharprime}{\kern0pt}\ {\isasymin}\ sets\ M{\isacharprime}{\kern0pt}\ {\isasymLongrightarrow}\ {\isacharparenleft}{\kern0pt}{\isasymlambda}{\isasymomega}{\isachardot}{\kern0pt}\ {\isasymkappa}\ {\isasymomega}\ A{\isacharprime}{\kern0pt}{\isacharparenright}{\kern0pt}\ {\isasymin}\ M\ {\isasymrightarrow}\isactrlsub M\ borel{\isachardoublequoteclose}\ \isakeyword{and}\isanewline
\ \ \ \ space{\isacharunderscore}{\kern0pt}source{\isacharunderscore}{\kern0pt}measure{\isacharcolon}{\kern0pt}\ {\isachardoublequoteopen}{\isasymAnd}{\isasymomega}{\isachardot}{\kern0pt}\ {\isasymomega}\ {\isasymin}\ space\ M\ {\isasymLongrightarrow}\isanewline
\ \ \ \ \ \ measure{\isacharunderscore}{\kern0pt}space\ {\isacharparenleft}{\kern0pt}space\ M{\isacharprime}{\kern0pt}{\isacharparenright}{\kern0pt}\ {\isacharparenleft}{\kern0pt}sets\ M{\isacharprime}{\kern0pt}{\isacharparenright}{\kern0pt}\ {\isacharparenleft}{\kern0pt}{\isasymlambda}A{\isacharprime}{\kern0pt}{\isachardot}{\kern0pt}\ {\isasymkappa}\ {\isasymomega}\ A{\isacharprime}{\kern0pt}{\isacharparenright}{\kern0pt}{\isachardoublequoteclose}%
}%
\DefineSnippet{kernel}{
\isacommand{typedef}\isamarkupfalse%
\ {\isacharparenleft}{\kern0pt}{\isacharprime}{\kern0pt}a{\isacharcomma}{\kern0pt}\ {\isacharprime}{\kern0pt}b{\isacharparenright}{\kern0pt}\ kernel\ {\isacharequal}{\kern0pt}\isanewline
\ \ {\isachardoublequoteopen}{\isacharbraceleft}{\kern0pt}{\isacharparenleft}{\kern0pt}M\ {\isacharcolon}{\kern0pt}{\isacharcolon}{\kern0pt}\ {\isacharprime}{\kern0pt}a\ measure{\isacharcomma}{\kern0pt}\ M{\isacharprime}{\kern0pt}\ {\isacharcolon}{\kern0pt}{\isacharcolon}{\kern0pt}\ {\isacharprime}{\kern0pt}b\ measure{\isacharcomma}{\kern0pt}\ {\isasymkappa}{\isacharparenright}{\kern0pt}{\isachardot}{\kern0pt}\ \isanewline
\ \ \ \ transition{\isacharunderscore}{\kern0pt}kernel\ M\ M{\isacharprime}{\kern0pt}\ {\isasymkappa}\ {\isasymand}\ {\isacharparenleft}{\kern0pt}{\isasymforall}{\isasymomega}{\isachardot}{\kern0pt}\ {\isasymforall}A{\isacharprime}{\kern0pt}{\isachardot}{\kern0pt}\ {\isasymomega}\ {\isasymin}\ {\isacharminus}{\kern0pt}space\ M\ {\isasymor}\ A{\isacharprime}{\kern0pt}\ {\isasymin}\ {\isacharminus}{\kern0pt}sets\ M{\isacharprime}{\kern0pt}\ {\isasymlongrightarrow}\ {\isasymkappa}\ {\isasymomega}\ A{\isacharprime}{\kern0pt}\ {\isacharequal}{\kern0pt}\ {\isadigit{0}}{\isacharparenright}{\kern0pt}{\isacharbraceright}{\kern0pt}{\isachardoublequoteclose}%
}%
\DefineSnippet{kernel_measure}{
\isacommand{lift{\isacharunderscore}{\kern0pt}definition}\isamarkupfalse%
\ kernel{\isacharunderscore}{\kern0pt}measure\ {\isacharcolon}{\kern0pt}{\isacharcolon}{\kern0pt}\ {\isachardoublequoteopen}{\isacharparenleft}{\kern0pt}{\isacharprime}{\kern0pt}a{\isacharcomma}{\kern0pt}\ {\isacharprime}{\kern0pt}b{\isacharparenright}{\kern0pt}\ kernel\ {\isasymRightarrow}\ {\isacharprime}{\kern0pt}a\ {\isasymRightarrow}\ {\isacharprime}{\kern0pt}b\ measure{\isachardoublequoteclose}\ \isakeyword{is}\isanewline
{\isachardoublequoteopen}{\isasymlambda}{\isacharparenleft}{\kern0pt}M{\isacharcomma}{\kern0pt}\ M{\isacharprime}{\kern0pt}{\isacharcomma}{\kern0pt}\ {\isasymkappa}{\isacharparenright}{\kern0pt}\ {\isasymomega}{\isachardot}{\kern0pt}\ measure{\isacharunderscore}{\kern0pt}of\ {\isacharparenleft}{\kern0pt}space\ M{\isacharprime}{\kern0pt}{\isacharparenright}{\kern0pt}\ {\isacharparenleft}{\kern0pt}sets\ M{\isacharprime}{\kern0pt}{\isacharparenright}{\kern0pt}\ {\isacharparenleft}{\kern0pt}{\isasymlambda}A{\isacharprime}{\kern0pt}{\isachardot}{\kern0pt}\ {\isasymkappa}\ {\isasymomega}\ A{\isacharprime}{\kern0pt}{\isacharparenright}{\kern0pt}{\isachardoublequoteclose}%
}%
\DefineSnippet{kernel_measure_pair_integral_measurable}{
\isacommand{lemma}\isamarkupfalse%
\ kernel{\isacharunderscore}{\kern0pt}measure{\isacharunderscore}{\kern0pt}pair{\isacharunderscore}{\kern0pt}integral{\isacharunderscore}{\kern0pt}measurable\ {\isacharbrackleft}{\kern0pt}measurable{\isacharbrackright}{\kern0pt}{\isacharcolon}{\kern0pt}\isanewline
\ \ \isakeyword{fixes}\ f\ {\isacharcolon}{\kern0pt}{\isacharcolon}{\kern0pt}\ {\isachardoublequoteopen}{\isacharprime}{\kern0pt}a\ {\isasymtimes}\ {\isacharprime}{\kern0pt}b\ {\isasymRightarrow}\ ennreal{\isachardoublequoteclose}\isanewline
\ \ \ \ \isakeyword{and}\ K\ {\isacharcolon}{\kern0pt}{\isacharcolon}{\kern0pt}\ {\isachardoublequoteopen}{\isacharparenleft}{\kern0pt}{\isacharprime}{\kern0pt}a{\isacharcomma}{\kern0pt}\ {\isacharprime}{\kern0pt}b{\isacharparenright}{\kern0pt}\ kernel{\isachardoublequoteclose}\isanewline
\ \ \isakeyword{assumes}\ f{\isacharunderscore}{\kern0pt}measurable{\isacharbrackleft}{\kern0pt}measurable{\isacharbrackright}{\kern0pt}{\isacharcolon}{\kern0pt}\ {\isachardoublequoteopen}f\ {\isasymin}\ borel{\isacharunderscore}{\kern0pt}measurable\ {\isacharparenleft}{\kern0pt}kernel{\isacharunderscore}{\kern0pt}source\ K\ {\isasymOtimes}\isactrlsub M\ kernel{\isacharunderscore}{\kern0pt}target\ K{\isacharparenright}{\kern0pt}{\isachardoublequoteclose}\isanewline
\ \ \ \ \ \ \isakeyword{and}\ finite{\isacharunderscore}{\kern0pt}K{\isacharcolon}{\kern0pt}\ {\isachardoublequoteopen}finite{\isacharunderscore}{\kern0pt}kernel\ K{\isachardoublequoteclose}\isanewline
\ \ \isakeyword{defines}\ {\isachardoublequoteopen}I\ {\isasymequiv}\ {\isasymlambda}\ f{\isachardot}{\kern0pt}\ {\isacharparenleft}{\kern0pt}{\isasymlambda}{\isasymomega}\isactrlsub {\isadigit{1}}{\isachardot}{\kern0pt}\ {\isasymintegral}\isactrlsup {\isacharplus}{\kern0pt}{\isasymomega}\isactrlsub {\isadigit{2}}{\isachardot}{\kern0pt}\ f{\isacharparenleft}{\kern0pt}{\isasymomega}\isactrlsub {\isadigit{1}}{\isacharcomma}{\kern0pt}\ {\isasymomega}\isactrlsub {\isadigit{2}}{\isacharparenright}{\kern0pt}\ {\isasympartial}kernel{\isacharunderscore}{\kern0pt}measure\ K\ {\isasymomega}\isactrlsub {\isadigit{1}}{\isacharparenright}{\kern0pt}{\isachardoublequoteclose}\isanewline
\ \ \isakeyword{shows}\ {\isachardoublequoteopen}I\ f\ {\isasymin}\ borel{\isacharunderscore}{\kern0pt}measurable\ {\isacharparenleft}{\kern0pt}kernel{\isacharunderscore}{\kern0pt}source\ K{\isacharparenright}{\kern0pt}{\isachardoublequoteclose}%
}%
\DefineSnippet{local_holder_on}{
\isacommand{definition}\isamarkupfalse%
\ local{\isacharunderscore}{\kern0pt}holder{\isacharunderscore}{\kern0pt}on\ {\isacharcolon}{\kern0pt}{\isacharcolon}{\kern0pt}\ {\isachardoublequoteopen}real\ {\isasymRightarrow}\ {\isacharprime}{\kern0pt}a\ {\isacharcolon}{\kern0pt}{\isacharcolon}{\kern0pt}\ metric{\isacharunderscore}{\kern0pt}space\ set\ {\isasymRightarrow}\ {\isacharparenleft}{\kern0pt}{\isacharprime}{\kern0pt}a\ {\isasymRightarrow}\ {\isacharprime}{\kern0pt}b\ {\isacharcolon}{\kern0pt}{\isacharcolon}{\kern0pt}\ metric{\isacharunderscore}{\kern0pt}space{\isacharparenright}{\kern0pt}\ {\isasymRightarrow}\ bool{\isachardoublequoteclose}\ \isakeyword{where}\isanewline
{\isachardoublequoteopen}local{\isacharunderscore}{\kern0pt}holder{\isacharunderscore}{\kern0pt}on\ {\isasymgamma}\ D\ {\isasymphi}\ {\isasymequiv}\ {\isasymgamma}\ {\isasymin}\ {\isacharbraceleft}{\kern0pt}{\isadigit{0}}{\isacharless}{\kern0pt}{\isachardot}{\kern0pt}{\isachardot}{\kern0pt}{\isadigit{1}}{\isacharbraceright}{\kern0pt}\ {\isasymand}\isanewline
\ \ {\isacharparenleft}{\kern0pt}{\isasymforall}t{\isasymin}D{\isachardot}{\kern0pt}\ {\isasymexists}{\isasymepsilon}\ {\isachargreater}{\kern0pt}\ {\isadigit{0}}{\isachardot}{\kern0pt}\ {\isasymexists}C\ {\isasymge}\ {\isadigit{0}}{\isachardot}{\kern0pt}\ {\isacharparenleft}{\kern0pt}{\isasymforall}r{\isasymin}D{\isachardot}{\kern0pt}\ {\isasymforall}s{\isasymin}D{\isachardot}{\kern0pt}\ dist\ s\ t\ {\isacharless}{\kern0pt}\ {\isasymepsilon}\ {\isasymand}\ dist\ r\ t\ {\isacharless}{\kern0pt}\ {\isasymepsilon}\ {\isasymlongrightarrow}\ dist\ {\isacharparenleft}{\kern0pt}{\isasymphi}\ r{\isacharparenright}{\kern0pt}\ {\isacharparenleft}{\kern0pt}{\isasymphi}\ s{\isacharparenright}{\kern0pt}\ {\isasymle}\ C\ {\isacharasterisk}{\kern0pt}\ dist\ r\ s\ powr\ {\isasymgamma}{\isacharparenright}{\kern0pt}{\isacharparenright}{\kern0pt}{\isachardoublequoteclose}%
}%
\DefineSnippet{holder_on}{
\isacommand{definition}\isamarkupfalse%
\ holder{\isacharunderscore}{\kern0pt}on\ {\isacharcolon}{\kern0pt}{\isacharcolon}{\kern0pt}\ {\isachardoublequoteopen}real\ {\isasymRightarrow}\ {\isacharprime}{\kern0pt}a\ {\isacharcolon}{\kern0pt}{\isacharcolon}{\kern0pt}\ metric{\isacharunderscore}{\kern0pt}space\ set\ {\isasymRightarrow}\ {\isacharparenleft}{\kern0pt}{\isacharprime}{\kern0pt}a\ {\isasymRightarrow}\ {\isacharprime}{\kern0pt}b\ {\isacharcolon}{\kern0pt}{\isacharcolon}{\kern0pt}\ metric{\isacharunderscore}{\kern0pt}space{\isacharparenright}{\kern0pt}\ {\isasymRightarrow}\ bool{\isachardoublequoteclose}\ {\isacharparenleft}{\kern0pt}{\isachardoublequoteopen}{\isacharunderscore}{\kern0pt}{\isacharminus}{\kern0pt}holder{\isacharprime}{\kern0pt}{\isacharunderscore}{\kern0pt}on{\isachardoublequoteclose}\ {\isadigit{1}}{\isadigit{0}}{\isadigit{0}}{\isadigit{0}}{\isacharparenright}{\kern0pt}\ \isakeyword{where}\isanewline
{\isachardoublequoteopen}{\isasymgamma}{\isacharminus}{\kern0pt}holder{\isacharunderscore}{\kern0pt}on\ D\ {\isasymphi}\ {\isasymlongleftrightarrow}\ {\isasymgamma}\ {\isasymin}\ {\isacharbraceleft}{\kern0pt}{\isadigit{0}}{\isacharless}{\kern0pt}{\isachardot}{\kern0pt}{\isachardot}{\kern0pt}{\isadigit{1}}{\isacharbraceright}{\kern0pt}\ {\isasymand}\ {\isacharparenleft}{\kern0pt}{\isasymexists}C\ {\isasymge}\ {\isadigit{0}}{\isachardot}{\kern0pt}\ {\isacharparenleft}{\kern0pt}{\isasymforall}r{\isasymin}D{\isachardot}{\kern0pt}\ {\isasymforall}s{\isasymin}D{\isachardot}{\kern0pt}\ dist\ {\isacharparenleft}{\kern0pt}{\isasymphi}\ r{\isacharparenright}{\kern0pt}\ {\isacharparenleft}{\kern0pt}{\isasymphi}\ s{\isacharparenright}{\kern0pt}\ {\isasymle}\ C\ {\isacharasterisk}{\kern0pt}\ dist\ r\ s\ powr\ {\isasymgamma}{\isacharparenright}{\kern0pt}{\isacharparenright}{\kern0pt}{\isachardoublequoteclose}%
}%
\DefineSnippet{local_holder_refine}{
\isacommand{lemma}\isamarkupfalse%
\ local{\isacharunderscore}{\kern0pt}holder{\isacharunderscore}{\kern0pt}refine{\isacharcolon}{\kern0pt}\isanewline
\ \ \isakeyword{assumes}\ g{\isacharcolon}{\kern0pt}\ {\isachardoublequoteopen}local{\isacharunderscore}{\kern0pt}holder{\isacharunderscore}{\kern0pt}on\ g\ D\ {\isasymphi}{\isachardoublequoteclose}\ {\isachardoublequoteopen}g\ {\isasymle}\ {\isadigit{1}}{\isachardoublequoteclose}\ \isanewline
\ \ \ \ \ \isakeyword{and}\ \ h{\isacharcolon}{\kern0pt}\ {\isachardoublequoteopen}h\ {\isasymle}\ g{\isachardoublequoteclose}\ {\isachardoublequoteopen}h\ {\isachargreater}{\kern0pt}\ {\isadigit{0}}{\isachardoublequoteclose}\isanewline
\ \ \ \isakeyword{shows}\ {\isachardoublequoteopen}local{\isacharunderscore}{\kern0pt}holder{\isacharunderscore}{\kern0pt}on\ h\ D\ {\isasymphi}{\isachardoublequoteclose}%
}%
\DefineSnippet{holder_uniform_continuous}{
\isacommand{lemma}\isamarkupfalse%
\ holder{\isacharunderscore}{\kern0pt}uniform{\isacharunderscore}{\kern0pt}continuous{\isacharcolon}{\kern0pt}\isanewline
\ \ \isakeyword{assumes}\ {\isachardoublequoteopen}{\isasymgamma}{\isacharminus}{\kern0pt}holder{\isacharunderscore}{\kern0pt}on\ X\ {\isasymphi}{\isachardoublequoteclose}\isanewline
\ \ \isakeyword{shows}\ {\isachardoublequoteopen}uniformly{\isacharunderscore}{\kern0pt}continuous{\isacharunderscore}{\kern0pt}on\ X\ {\isasymphi}{\isachardoublequoteclose}%
}%
\DefineSnippet{holder_implies_local_holder}{
\isacommand{lemma}\isamarkupfalse%
\ holder{\isacharunderscore}{\kern0pt}implies{\isacharunderscore}{\kern0pt}local{\isacharunderscore}{\kern0pt}holder{\isacharcolon}{\kern0pt}\ {\isachardoublequoteopen}{\isasymgamma}{\isacharminus}{\kern0pt}holder{\isacharunderscore}{\kern0pt}on\ D\ {\isasymphi}\ {\isasymLongrightarrow}\ local{\isacharunderscore}{\kern0pt}holder{\isacharunderscore}{\kern0pt}on\ {\isasymgamma}\ D\ {\isasymphi}{\isachardoublequoteclose}%
}%
\DefineSnippet{local_holder_compact_imp_holder}{
\isacommand{lemma}\isamarkupfalse%
\ local{\isacharunderscore}{\kern0pt}holder{\isacharunderscore}{\kern0pt}compact{\isacharunderscore}{\kern0pt}imp{\isacharunderscore}{\kern0pt}holder{\isacharcolon}{\kern0pt}\isanewline
\ \ \isakeyword{assumes}\ {\isachardoublequoteopen}compact\ I{\isachardoublequoteclose}\ {\isachardoublequoteopen}local{\isacharunderscore}{\kern0pt}holder{\isacharunderscore}{\kern0pt}on\ {\isasymgamma}\ I\ {\isasymphi}{\isachardoublequoteclose}\isanewline
\ \ \isakeyword{shows}\ {\isachardoublequoteopen}{\isasymgamma}{\isacharminus}{\kern0pt}holder{\isacharunderscore}{\kern0pt}on\ I\ {\isasymphi}{\isachardoublequoteclose}%
}%
\DefineSnippet{nn_integral_Markov_inequality_extended}{
\isacommand{lemma}\isamarkupfalse%
\ nn{\isacharunderscore}{\kern0pt}integral{\isacharunderscore}{\kern0pt}Markov{\isacharunderscore}{\kern0pt}inequality{\isacharunderscore}{\kern0pt}extended{\isacharcolon}{\kern0pt}\isanewline
\ \ \isakeyword{fixes}\ f\ {\isacharcolon}{\kern0pt}{\isacharcolon}{\kern0pt}\ {\isachardoublequoteopen}real\ {\isasymRightarrow}\ ennreal{\isachardoublequoteclose}\ \isakeyword{and}\ {\isasymepsilon}\ {\isacharcolon}{\kern0pt}{\isacharcolon}{\kern0pt}\ real\ \isakeyword{and}\ X\ {\isacharcolon}{\kern0pt}{\isacharcolon}{\kern0pt}\ {\isachardoublequoteopen}{\isacharprime}{\kern0pt}a\ {\isasymRightarrow}\ real{\isachardoublequoteclose}\isanewline
\ \ \isakeyword{assumes}\ mono{\isacharcolon}{\kern0pt}\ {\isachardoublequoteopen}mono{\isacharunderscore}{\kern0pt}on\ {\isacharparenleft}{\kern0pt}range\ X\ {\isasymunion}\ {\isacharbraceleft}{\kern0pt}{\isadigit{0}}{\isacharless}{\kern0pt}{\isachardot}{\kern0pt}{\isachardot}{\kern0pt}{\isacharbraceright}{\kern0pt}{\isacharparenright}{\kern0pt}\ f{\isachardoublequoteclose}\isanewline
\ \ \ \ \ \ \isakeyword{and}\ finite{\isacharcolon}{\kern0pt}\ {\isachardoublequoteopen}{\isasymAnd}x{\isachardot}{\kern0pt}\ f\ x\ {\isacharless}{\kern0pt}\ {\isasyminfinity}{\isachardoublequoteclose}\isanewline
\ \ \ \ \ \ \isakeyword{and}\ e{\isacharcolon}{\kern0pt}\ {\isachardoublequoteopen}{\isasymepsilon}\ {\isachargreater}{\kern0pt}\ {\isadigit{0}}{\isachardoublequoteclose}\ {\isachardoublequoteopen}f\ {\isasymepsilon}\ {\isachargreater}{\kern0pt}\ {\isadigit{0}}{\isachardoublequoteclose}\isanewline
\ \ \ \ \ \ \isakeyword{and}\ {\isacharbrackleft}{\kern0pt}measurable{\isacharbrackright}{\kern0pt}{\isacharcolon}{\kern0pt}\ {\isachardoublequoteopen}X\ {\isasymin}\ borel{\isacharunderscore}{\kern0pt}measurable\ M{\isachardoublequoteclose}\isanewline
\ \ \ \ \isakeyword{shows}\ {\isachardoublequoteopen}emeasure\ M\ {\isacharbraceleft}{\kern0pt}p\ {\isasymin}\ space\ M{\isachardot}{\kern0pt}\ {\isacharparenleft}{\kern0pt}X\ p{\isacharparenright}{\kern0pt}\ {\isasymge}\ {\isasymepsilon}{\isacharbraceright}{\kern0pt}\ {\isasymle}\ {\isacharparenleft}{\kern0pt}{\isasymintegral}\isactrlsup {\isacharplus}{\kern0pt}\ x{\isachardot}{\kern0pt}\ f\ {\isacharparenleft}{\kern0pt}X\ x{\isacharparenright}{\kern0pt}\ {\isasympartial}M{\isacharparenright}{\kern0pt}\ {\isacharslash}{\kern0pt}\ f\ {\isasymepsilon}{\isachardoublequoteclose}%
}%

\title{Brownian motion in Isabelle/HOL}
\author{Christian Pardillo Laursen \and Simon Foster \and Mark Post}
\institute{University of York}

\begin{document}
\maketitle

\begin{abstract}
In order to formally verify robotic controllers, we must tackle the inherent uncertainty of sensing and actuation in a physical environment. We can model uncertainty using stochastic hybrid systems, which combine discrete jumps with continuous, stochastic behaviour. The verification of these systems is intractable for state-exploration based approaches, so we instead propose a deductive verification approach. As a first step towards a deductive verification tool, we present a mechanisation of Brownian motion within Isabelle/HOL. For this, we mechanise stochastic kernels and Markov semigroups, which allow us to specify a range of processes with stationary, independent increments. Further, we prove the Kolmogorov-Chentsov theorem, which allows us to construct H{\"o}lder continuous modifications of processes that satisfy certain bounds on their expectation. This paves the way for modelling and verifying stochastic hybrid systems in Isabelle/HOL.\end{abstract}

\section{Introduction}

Formal verification of robotic systems is a difficult problem. Robots are systems operating under uncertainty in continuous time and continuous space, which have to be taken into consideration when verifying their control software. We can capture this uncertainty by modelling robots as stochastic hybrid systems (SHSs), which combine stochastic dynamics with discrete state transitions~\cite{pola03}.

However, there are challenges associated with formally verifying SHSs. Automated approaches generally require discretising the state space~\cite{abate10, david12}. This makes the verification less accurate and often the state space is too large after discretisation, especially considering the stochastic component. To overcome these limitations, we propose applying deductive verification to SHSs.

The main advantage of this approach is that we can reason symbolically, obviating the need for discretisation and guiding the verification through human input. Deductive verification requires more expert input from its users than automated approaches, but it does not require approximations and is not restricted to linear SDEs, making this approach more widely applicable and its verification guarantees stronger. Several authors have proposed proof calculi for deductively verifying SHS~\cite{platzer11, wang17}, but as of yet, none have been implemented.

Our vision is a novel approach for the modelling and verification of SHS based on the interactive proof assistant Isabelle/HOL~\cite{isabelle}. Isabelle/HOL is a general-purpose proof assistant for higher-order logic, and its theorems are built from first principles in typed set theory. This makes any logic developed within Isabelle/HOL inherently sound, modular, and conservatively extensible. We can also harness established mechanised theory libraries, particularly within probability theory~\cite{hirata23, keskin23}, and automated reasoning techniques already implemented for Isabelle~\cite{sledgehammer, munive24}, making our results more extensible and wide-reaching.

This approach has been successfully applied in the IsaVODEs tool~\cite{foster21,munive24}. The authors used the multivariate analysis library~\cite{Harrison2005-Euclidean} of Isabelle/HOL as the foundation for implementing differential induction, which is the backbone of differential dynamic logic~\cite{platzer08}. This results in a powerful, fully formalised implementation of a hybrid systems verification logic. We propose to follow in their footsteps by adapting and implementing Platzer's proposed logic for SHS, stochastic differential dynamic logic~\cite{platzer11}, which introduces a probabilistic diamond modality that lets users reason about stochastic evolution in continuous time.

As a first step towards this, we mechanise Brownian motion in Isabelle/HOL. This is an important continuous-time stochastic process which is used to give semantics to stochastic differential equations. This in turn describes the dynamics of SHS, forming the foundation of our proposed logic. Brownian motion is a central object in probability theory --- it models the random motion of particles in a fluid, but it can be applied to model a wide range of phenomena, such as thermal properties~\cite{jang04} and stock prices~\cite{osborne59}.

We identify three key contributions in this work. Firstly, we develop a method for constructing stochastic processes from families of stochastic kernels \-- stochastic transition systems which generalise Markov chains. Secondly, we formalise the Kolmogorov-Chentsov theorem which enables the construction of a process with almost surely continuous paths. Finally, using these two results, we construct Brownian motion. All of our results have been formalised in Isabelle/HOL\footnote{The results are available at \url{https://github.com/cplaursen/Brownian_Motion}}, and mainly follow Klenke's ``Probability theory: A Comprehensive Course'' text; one of the key textbooks in probability theory~\cite{klenke20}.

Our main contribution is not the theory of stochastic processes itself, but instead the engineering effort involved in formalising it. The mechanisation process goes beyond reiterating the results: pen-and-paper proofs are always laden with omissions, and their mechanisation strengthens their argument by spelling out every detail of the proof. This often involves creativity, especially in the case where the textbook definitions or problem statements have to be altered to fit within the logic or enable better automation. We also gain insights into mechanisation of non-trivial continuous mathematics, and how proof automation both aids development and falls flat. In our development, we benefited from Isabelle's proof automation -- both automated theorem proving with Sledgehammer~\cite{Blanchette2016Hammers}, but also more traditional and bespoke proof methods.

These results have not been mechanised before to the authors' knowledge, and they comprise of some deep -- yet fundamental -- results in stochastic process theory. Our mechanisation has required us to formalise a number of smaller novel technical results, particularly the iterated kernel product and a constructive definition for intervals of dyadic rationals. The results are also applicable outside of constructing Brownian motion; transition kernels can be used to give semantics to probabilistic programs~\cite{hirata23}, and the Kolmogorov-Chentsov theorem is applicable for constructing an important class of stochastic processes.

In \S\ref{sec:sim} we describe the modelling paradigm that we are aiming to formally verify using Isabelle/HOL. In \S\ref{sec:background} we give the required mathematical background for the content presented in this paper. We present our formalisation of stochastic processes in \S\ref{sec:process}, and then introduce transition kernels in \S\ref{sec:kernel}, which we will use to construct processes. We detail our formalisation of the Kolmogorov-Chentsov theorem and the construction of Brownian motion in \S\ref{sec:modification}. We then evaluate our contribution in \S\ref{sec:evaluation}, give an account of related work in \S\ref{sec:related-work} and summarise our contributions in \S\ref{sec:conclusion}.

\section{Modelling stochastic hybrid systems}\label{sec:sim}
\begin{figure}[t]
\centering
\tikzset{elliptic state/.style={draw, ellipse}}
\begin{tikzpicture}
    \node (off) [elliptic state,
        initial,
        initial left,
        initial distance=.5cm,
        thick,
        initial text=$t\eq20$] {$\begin{gathered}\textbf{Off}\\\dot{t} = -0.1t + 0.1 dB\\t\geq18\end{gathered}$};
    \node (on) [elliptic state] at (5,0) {$\begin{gathered}\textbf{On}\\\dot{t} = 5-0.1t + 0.1 dB\\t\leq22\end{gathered}$};
    \path [-stealth, thick]
    (off) edge[bend left] node [above]{$t<19$}   (on)
    (on) edge[bend left] node [below]{$t>21$} (off);
\end{tikzpicture}
\includegraphics[width=0.9\textwidth]{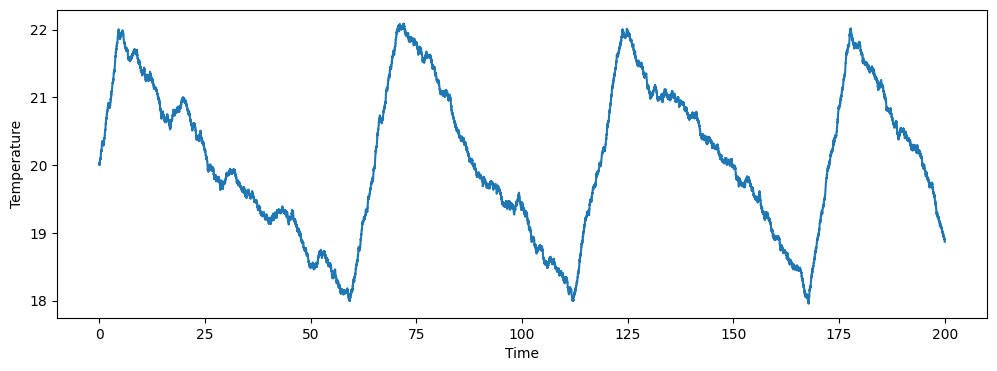}

\vspace{-2ex}
\caption{Thermostat hybrid automaton and simulation, adapted from Henzinger~\cite{henzinger00}}\label{fig:thermostat}

\vspace{-1ex}
\end{figure}
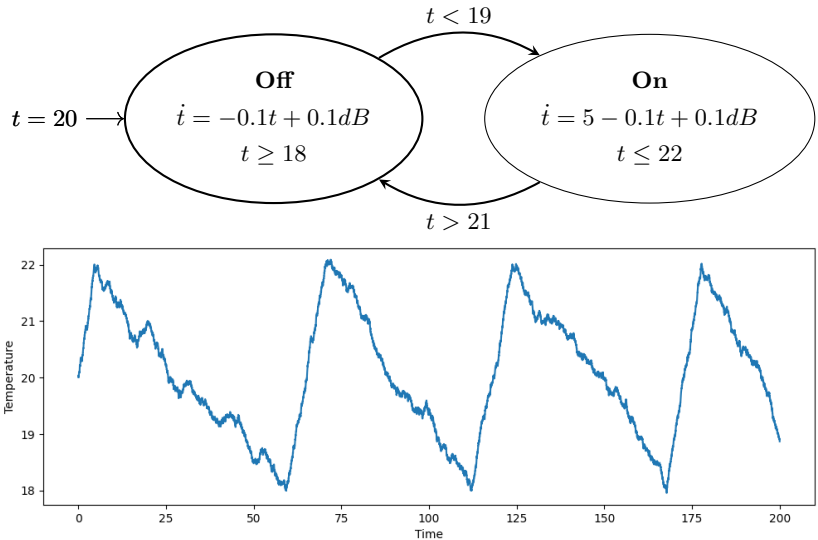

There are a few competing definitions of SHS, as documented by Pola et al.~\cite{pola03}. The most adequate for our purposes are the Stochastic Hybrid Systems described by Hu et al.~\cite{hu00} --- state machines where each discrete state is associated with a stochastic differential equation, which specifies how the continuous state evolves. The system evolves continuously until it reaches a boundary condition, at which point a discrete state transition happens instantaneously. We exemplify this modelling paradigm using the stochastic thermostat presented in Figure~\ref{fig:thermostat}.

This is a thermostat operating in a stochastic environment, displayed in the style of hybrid automata~\cite{henzinger00}. It has two states, on and off, which it switches between when the jump conditions are met. The continuous evolution of the temperature is given by stochastic differential equations, which combine deterministic drift with a stochastic noise component. $dB$ refers to the differential of the Brownian motion process. This paradigm allows us to represent systems that experience stochastic interference since we can model the noise directly. In our example, temperature may fluctuate randomly but it will tend to increase when the heating is on, and slowly decrease when it is turned off.

While simple, the behaviour of this model reflects the temperature fluctuations that happen in the real world, which cannot accurately be modelled deterministically. This uncertainty in the evolution of the continuous system appears in various forms in many applications, and being able to incorporate it into our models directly -- instead of abstracting it away -- shrinks the reality gap~\cite{jakobi95}.

\section{Mathematical background}\label{sec:background}
Our development is based on the Isabelle/HOL analysis and probability libraries, namely \textsf{HOL-Analysis} and \textsf{HOL-Probability}~\cite{holzl11, holzlphd}. Here, we recount some notions from probability and measure theory that we shall use throughout the paper.

The central objects of measure theory are measure spaces: triples \((\Omega, \mathcal{F}, \mu)\) consisting of the space \(\Omega\), the measurable sets \(\mathcal{F}\) and a measure function \(\mu : \mathcal{F} \to \mathbb{R}\). The measurable sets form a $\sigma$-algebra: a family of sets closed under countable unions and complements, which contains $\Omega$. A probability space is a measure space for which \(\mu(\Omega) = 1\).

Given measurable spaces \((X, \mathcal{F})\) and \((Y, \mathcal{G})\), \(f : X \to Y\) is \(\mathcal{F} - \mathcal{G}\) measurable if for any \(S \in \mathcal{G}\), \(\{x \mid f(x) \in S\} \in \mathcal{F}\). In the setting of probability theory, measurable functions are called random variables. When operating on random variables, it is standard practice to omit the points of the source measure: we write \(\{X \neq Y\}\) instead of \(\{\omega \in \Omega. X \omega \neq Y \omega\}\). The distribution of a measurable function is the push-forward measure they induce on their target, defined by \(\mathcal{D}_X(S) = \mu(\{\omega \in \Omega \mid X(\omega) \in S\})\).

Exclusive to probability is the notion of independence: two families of events $\mathcal{A}$ and $\mathcal{B}$ are independent with respect to probability measure \(\mu\) if \(\forall A \in \mathcal{A}, B \in \mathcal{B}.\, \mu(A)\cdot\mu(B) = \mu(A \cap B)\). Then, random variables are independent if their preimages form independent families. We state that a property $P$ holds almost everywhere (a.e.), or almost surely (a.s.) in a probability space, if \(\{x \mid \lnot P(x)\}\) is a subset of a measurable set with measure 0 (a null set).

The product measure \(\prod_{i\in I} (\Omega_i, \mathcal{F}_i, \mu_i)\) is generated by sets of the form \(\{A \mid \forall i \in I. A_i \in \mathcal{F}_i\}\). The variable \(X_J\) for \(J \subseteq I\) is the projection from the product space with index \(I\) to index \(J\) - this helps us characterise product spaces with infinite index sets by considering their finite-dimensional distributions.

The Lebesgue integral generalises the Riemann integral by allowing us to integrate any measurable function, and to do so with respect to any measure.

\section{Stochastic processes}\label{sec:process}
In this section, we introduce stochastic processes, modifications and indistinguishability, and convergence of processes in measure, all of which will be necessary to prove the existence of Brownian motion.

A stochastic process is an indexed family of random variables from some probability space \(\{X_t \mid t \in I\}\) for random variables \(X_t\) and index set $I$. In many instances $I$ represents time; a common choice is the natural numbers or the interval \([0, \infty)\). An example is the Bernoulli process, a discrete time stochastic process with $I = \mathbb{N}$, which has two outcomes, 0 and 1, and can be used to model a sequence of coin tosses. Another example is a one dimensional simple random walk, where in each time-step a position is incremented or decremented by 1, based on the outcome of the Bernoulli process. We define stochastic processes in Isabelle using a locale:

\Snippet{stochastic_process}

\noindent Locales allow us to implement algebraic structures, including their carrier sets, operations and axioms. A locale is a named set of fixed variables and assumptions~\cite{haftmann09}. One can prove theorems within a locale relative to the locale assumptions and fixed variables, whilst making use of other theorems in the locale.

Our stochastic process locale extends \texttt{prob\_space}, another locale which fixes \texttt{M :: 'a measure}, i.e. a measure space taking values of type \texttt{'a}, and assumes it is a probability space. To these assumptions we add three fixed variables: \(M'\) is the target measurable space of our process, \(I\) is the index set, and $X$ is the process itself. We then assume that that \(X_i\) is a random variable into space \(M'\) for all \(i\). We add the \texttt{[measurable]} theorem attribute to our assumption, which tags it as part of the named theorem set \texttt{measurable} and enables the use of the measurability tactic, which will use this fact to prove subgoals related to measurability of our process.

We can define the simple random walk in Isabelle by taking the source probability space to be the stream space on Booleans as the process $R$:
\[R_n(\omega) \triangleq \,\sum_{j\in\{1..n\}}\text{if }\omega!!j\text{ then }1\text{ else }-1\]
where $\omega !! j$ is a function that takes the $j$th element of stream $\omega$, which intuitively corresponds to the $j$th coin toss in a random sequence.

Locales define a predicate that characterises their fixed variables and assumptions. We use the predicate defined by the stochastic process locale to define the type of stochastic processes as a quadruple of source measure, target measure, index set and family of random variables, which satisfy the locale predicate.

We define the distributions of the process - given some \(J \subseteq I\), we define a product measure where the distribution at each index is given by the process:

\[X_J = \prod_{j \in J} \mathcal{D}_{X_j}\]

A process has independent increments if for all \(t_0,\dots,t_n\) with \(t_0 < t_1 < \dots < t_n\), all \((X_{t_i} - X_{t_{i-1}})_{i=1\dots n}\) are mutually independent. In Isabelle, we define increments as sorted lists: a suitable stochastic process has independent increments if for any sorted list of a least two indices drawn from index set I, every successive pair of variables are independent.

Using these definitions, we can characterise Brownian motion, also known as the Wiener process, as a process \(W\) which satisfies \(W_0 = 0\), has independent, normally distributed increments, and almost surely continuous paths. This process is an ideal starting point for modelling real-world systems exhibiting uncertainty, as it arises naturally in many kinds of noise. It is also very well-behaved. Independent increments mean that previous events do not influence future outcomes. Normally distributed increments are well-behaved and appropriate for many situations because of the central limit theorem. Finally, its continuous paths make it suitable for modelling physical phenomena, which cannot have discontinuities.

In Isabelle, we define Brownian motion using a locale:
\Snippet{brownian_motion}
\noindent This locale definition fixes a stochastic process $W$ over the reals, and asserts the four key properties introduced above.

During the rest of this text, we develop the machinery required for proving the existence of this process. On one hand, we will construct a process with independent, normally distributed increments by defining a family of probability distributions and making use of the Daniell-Kolmogorov theorem, available in Isabelle~\cite{immler12}, for extending this family to an uncountable product space. However, this does not guarantee almost surely continuous paths; for this we will prove the Kolmogorov-Chentsov theorem, which will allow us to modify a process in order to make its paths almost surely continuous.

\subsection{Comparing processes}
Often in probability theory we extend usual properties so that they hold outside of some set with probability 0. This is the case for equality between stochastic processes, and we present two different ways of extending it in this way.

In order to compare processes, we first introduce the concept of compatible processes, which share the same index set, target measure, and source $\sigma$-algebra. Any two compatible processes $X$ and $Y$ are modifications of each other if, for any given $t \in I$, $P[X_t = Y_t] = 1$ Equivalently, $X$ is a modification of $Y$ iff there exists a family of null sets $N_t$ such that $\{X_t \neq Y_t\} \subseteq N_t$ for all $t \in I$.

We present the following proof about modifications from our development as a sample of what the Isabelle mechanisation process entails:
\Snippet{modification_sym}
\noindent The proof consists of unpacking our definitions of modification and compatibility using the destructor laws we define in order to make use of the congruence laws of almost everywhere and the symmetry of equality.

Two processes $X$ and $Y$ are indistinguishable if the set $\{\exists t.\, X t \neq Y t\}$ is indeed a null set. Indistinguishability clearly implies modification, and we prove that the other direction of implication holds for a countable index set, and almost surely continuous processes on an interval:

\begin{lemma}
Let $X$ and $Y$ be stochastic processes into a metric space \((S, d)\) which are modifications of each other, and that their index takes the form \([0,T]\). Assume that $X$ and $Y$ are almost surely continuous. Then $X$ and $Y$ are indistinguishable.
\end{lemma}

The proof of this theorem relies on obtaining a null set for a countable dense subset of the index space, and extending the null set to cover the whole space by continuity. We also rely on this technique later for the proof of the Kolmogorov-Chentsov theorem. For this use the dyadic rationals: numbers of the form \(\frac{k}{2^n}\) where \(k, n\in\mathbb{N}\). We present a novel definition for the sets of dyadic rationals with denominator \(2^n\) covering intervals of reals \([0,T]\) as follows:
 \[D_n(T) = \left\{\frac{k}{2^n} \mid k \in 0 \ldots\left\lfloor 2^nT \right\rfloor\right\}\]
Then, the dyadic rationals are defined by \(D(T)= \bigcup_n D_n(T)\). We prove that they are dense in the reals and thus satisfy the requirements for uniquely extending a uniformly continuous function defined on the dyadics to the reals.

\subsection{Convergence in measure}
The usual notion of convergence for functions is pointwise convergence: a sequence of functions \(f_i: X \to Y\) converges pointwise to \(f\) if \(\forall x \in X.\, \lim_{n \to \infty} f_n(x) = f(x)\). As in the previous section, there are two ways of generalising this within a measure space, where we want to consider convergence up to measure 0: convergence in measure and almost sure convergence.

The texts we consulted only define convergence in measure for sequences, and leave it to the reader to extrapolate for real-valued functions. We mechanise this by defining convergence in measure in terms of filters. These are commonly used in topology, and particularly Isabelle/HOL, to indicate the ``mode of convergence'' that we are interested in. Examples of filters are the convergence of a sequence as $n \to \infty$, or the limit of the function at a point $\lim_{x \to a}f(x)$. We write \(f \longrightarrow_F l\) to mean that function f converges to limit l along filter F.

A family of measurable functions $f$ converges in measure $\mu$ to measurable function $l$ in filter $F$ if for all measurable sets $A$ of finite measure and $\varepsilon > 0$:
\[\mu (\{\omega. d (f_n(\omega),l(\omega)) > \varepsilon\} \cap A)) \longrightarrow_F 0)\]

A family of measurable functions $f$ converges almost everywhere to measurable function $l$ in filter $F$ if
\[f_n \longrightarrow_F l \qquad \text{almost everywhere}\]

By intersecting with the set of finite measure in our definition of convergence in measure, we can show that convergence almost everywhere implies convergence in measure, which are both weaker notions than standard pointwise convergence. In finite measure spaces, the definition of convergence in measure is equivalent to one without the intersection with the set of finite measure. We show that the limit in measure -- and by consequence the limit almost everywhere -- is unique almost everywhere.

\section{Transition kernels}\label{sec:kernel}
Transition kernels can be understood as generalisations of Markov chain transition functions. They can be used to characterise and construct Markov processes in a natural way, and we will use them to construct our Brownian motion process. In this section, we develop the machinery required for defining stochastic processes using well-behaved families of kernels. This involves defining the infinite product space generated by a family of kernels, building on the Daniell-Kolmogorov theorem, whose coordinate process will have the desired properties. In particular, we are concerned with Markov semigroups: consistent families of kernels where composing kernels is equivalent to summing their indices.

We define a transition kernel from measurable space \((\Omega, \mathcal{A})\) to \((\Omega', \mathcal{A}')\) as a function \(\kappa: \Omega \times \mathcal{A}' \to \mathbb{R}^+\) such that:
\begin{align*}
\forall A' \in \mathcal{A}'.\quad& \kappa(-, A') \text{ is } \mathcal{A} - \mathcal{B}\text{ measurable}\\
\forall \omega \in \Omega.\quad& \kappa(\omega, -) \text{ is a measure on } (\Omega', \mathcal{A}')
\end{align*}

\noindent Intuitively, given some current state \(\omega\), \(\kappa(\omega, A')\) computes the probability of transitioning to some state in \(A'\). Transition kernels provide a natural way of defining processes by characterising their increments. For example, we can characterise a random walk on the integers that increases with probability $p$ and decreases with probability ($1-p$) by the following kernel on the integers:

\[\kappa(i, \{m\}) = \begin{cases}p&\mbox{ if $i = m-1$ } \\
1-p&\mbox{ if $i = m+1$ }
\\0&\mbox{ otherwise }
\end{cases}\]

\noindent Because we are working with a discrete sigma algebra, i.e. the power set of the integers, the kernel is fully defined by its operation on singleton sets. We make use of countable additivity to compute the probability of some event, by simply taking the sum over its elements.

In Isabelle, we define kernels as a locale:
\Snippet{transition_kernel}

A transition kernel \(\kappa : \Omega \to A'\) is stochastic iff \(\forall \omega \in \Omega.\, \kappa(\omega, -)\) is a probability measure. In order to compose kernels together we need to integrate with respect to the kernel measure. We show that this integral is measurable, which is essential if we are composing several integrals, or defining a kernel in terms of the integral of another kernel.
\begin{lemma}
    Let $f$ be an \((\Omega_1 \times \Omega_2 - \mathcal{B})\) measurable function. Let \(\kappa\) be a finite kernel on \(\Omega_1 - \Omega_2\). Define \[I_f(\omega_1) = \int f(\omega_1, \omega_2) \kappa(\omega_1, d\omega_2)\] Then \(I_f\) is a Borel measurable function.
\end{lemma}

The proof for this lemma follows an approximation approach for Borel measurable functions, where we first prove that the theorem holds for indicator functions, then simple functions, and finally for any Borel measurable functions as they can be represented as countable sums of simple functions, which are measurable because of the monotone convergence of the Lebesgue integral.

\subsection{Kernel Product}
The binary kernel product combines two kernels to operate on a product space. Let \(\kappa_1\) be an \(\Omega_1 - \Omega_2\) kernel, and \(\kappa_2\) be a \(\Omega_1 \times \Omega_2 - \Omega_3\) kernel. Then the kernel product \(\otimes_K\) is a kernel \(\Omega_1 - \Omega_2 \times \Omega_3\) defined by
\[(\kappa_1 \otimes_K \kappa_2)(\omega_0, A) \triangleq \int\left(\int \mathbb{1}_A(\omega_1, \omega_2)\kappa_2((\omega_0, \omega_1)d\omega_2)\right)\kappa_1(\omega_0, d\omega_1)\]

We also define \(\otimes_P\) as the kernel product of \(\kappa_1: \Omega_1 - \Omega_2\) and \(\kappa_2: \Omega_2 - \Omega_3\) by understanding \(\kappa_2\) as a kernel that ignores the elements of \(\Omega_1\). This operation has more reasonable types which makes it more compositional, and therefore more useful in defining stochastic processes.

We show that our definitions form a transition kernel if the kernels are finite. For this, we need to prove that the inside of the integral is measurable, which requires an additional proof. The arguments for these are quite standard - approximation using Dynkin systems. If both kernels are finite then the product is \(\sigma\)-finite, and if they are stochastic or substochastic then the product is too.

Our objective is to define the finite product kernel analogously to the following recursion:
\[\bigotimes_{i=1}^{k+1} \kappa_i = (\kappa_1 \otimes \kappa_2 \otimes \dots \otimes \kappa_{k}) \otimes \kappa_{k+1}\]
This is not possible within the type system of Isabelle, as it is not possible to construct a type of $n$-fold Cartesian products where $n$ is allowed to vary. Instead, we represent these $n$-fold products as functions from a finite index set, and define the recursion using the product measure \texttt{PiM} as our target.

Let $n$ be some natural number and $K$ a family of kernels with index set $\{0..n\}$. Then \(\bigotimes_{k=0}^n K_k\) is a kernel \(\Omega - \Omega^{\{0..n\}}\). We define this by recursion: the base case lifts kernel $K_0$ to operate on the product space of functions, and the recursive case defines the product kernel as an integral with respect to $K (n+1)$, analogously to our binary product construction. We show that this kernel is well-formed by induction on $n$, and we are able to reuse proofs from our binary product construction.

We use the product kernel to define a construction using convolution \(\star\), i.e. the distribution of \(+\):
\begin{lemma}\label{lemma:convolution}
Given a family of independent variables \(X_i\) and index I, define the kernels \[\kappa_i(\omega, -) \triangleq \delta_\omega \star \mathcal{D}_{X_i}\] The product kernel of \(\kappa_i\) at state 0 is equal to the distribution of \(X_i\):
\[\left(\bigotimes_{k=0}^n \kappa_{I_k}\right) 0 = \prod_{t\in I([0,n])}\mathcal{D}(X_t)\]
\end{lemma}

These product kernels are particularly well-behaved, and they will allow us to define a class of well-behaved stochastic processes.

We now define a way of obtaining a measure from a kernel. Let \((\Omega, \mathcal{F}, \mu)\) be a measure space, and let \(\kappa: \mathcal{F} - \mathcal{G}\) be a finite kernel. Then
\[\mu \otimes_S \kappa \triangleq \kappa_\mu \otimes_P \kappa\]
defines the semidirect product. We prove that integrals over the semidirect product can be split into two integrals:

\[\int f(\omega) (\mu \otimes_S \kappa)(d \omega) = \int \left(\int f(\omega_1, \omega_2) \kappa(\omega_1, d\omega_2)\right) \mu(d\omega_1)\]

Finally, we define kernel composition. Let \(\kappa_1\) be a finite \((\Omega_0, \mathcal{A}_0) - (\Omega_1, \mathcal{A}_1)\) transition kernel, and let \(\kappa_2\) be a \((\Omega_1, \mathcal{A}_1) - (\Omega_2, \mathcal{A}_2)\) transition kernel. We can now define kernel composition by \[\kappa_1 \circ \kappa_2(\omega_0, A_2) = \int_{\Omega_1} \kappa_2(\omega_1, A_2) \kappa_1(\omega_0, d\omega_1) \]

\subsection{Markov semigroups}
Let \(\kappa_{i,j} :I \to I \to \Omega \to \mathcal{A}\) be a family of stochastic kernels on \((\Omega, \mathcal{A})\), where $I$ is an ordered set. \(\kappa\) is a consistent family if:
\[\kappa_{i,j} \circ \kappa_{j,k} = \kappa_{i,k} \qquad \text{ for } i,j,k \in I\]

Using the finite kernel product, we show that consistent families of kernels form a projective family, and therefore we can construct an infinite product kernel using them.

\begin{lemma}
For any consistent family of stochastic kernels \(E - E\) with index set \(I \subseteq [0, \infty)\) containing 0, we show there is a kernel \(E - E^I\) such that, for all sets \(J \subseteq I = \{j_1, j_2, \dots, j_n\}\) where \(j_1 < j_2 < \dots < j_n\): 
    \[\kappa(x, -) \circ X_J^{-1} = \left(\delta_x \otimes_S \bigotimes_{k=0}^{n-1} \kappa_{j_k, j_{k+1}}\right)\]
\end{lemma}

Now consider the stochastic family of kernels \(\kappa_i : I \to \Omega\to \mathcal{A}\) on \((\Omega, \mathcal{A})\). \(\kappa_i\) is a Markov semigroup if
\begin{align*}
    \kappa_0(\omega, -) &= \delta_\omega\\
    \kappa_s \circ \kappa_t &= \kappa_{s+t}
\end{align*}

Any Markov semigroup \(\kappa\) is a consistent family defined by \(\kappa_{s,t} = \kappa_{t-s}\). The infinite product kernel defined by a Markov semigroup is particularly well-behaved, and defines a process with stationary, independent increments.

Using this, we can construct a stochastic process with independent, stationary increments for any family of distributions that satisfy the property \(\nu_{t+s} = \nu_t \star \nu_s\) using Lemma~\ref{lemma:convolution}.

\section{Kolmogorov-Chentsov}\label{sec:modification}
The Kolmogorov-Chentsov theorem allows us to construct continuous modifications of processes which satisfy certain conditions on their expectations. We first present some concepts needed for the statement and proof of the theorem.

\subsection{H{\"o}lder continuity}
Let \((E, d)\) and \((E', d')\) be two metric spaces. A function \(\varphi: E \to E'\) is locally H{\"o}lder continuous of order \(\gamma\) (H{\"o}lder-\(\gamma\)-continuous) for \(0 < \gamma \le 1\) on set \(D\) if for all \(t \in D\) there is an \(\varepsilon > 0\) and a \(C \ge 0\) such that
\[\forall r, s \in B_t(\varepsilon)\cap D.\, d(\varphi(r), \varphi(s)) \le  C d'(r,s)^\gamma\]

Where \(B_t\) is the open ball centred around t. \(\varphi\) is \(\gamma\)-(globally) H{\"o}lder continuous on $D$ if we don't restrict r and s to be in some neighbourhood of t - that is, if there exists some \(C \ge 0\) s.t.
\[\forall r, s \in D.\, d(\varphi(r), \varphi(s)) \ge C d'(r,s)^\gamma\]

Clearly, if \(\varphi\) is H{\"o}lder-\(\gamma\)-continuous on D it is also locally H{\"o}lder-\(\gamma\)-continuous on $D$. The other direction holds if $D$ is compact - we prove this by constructing the constant $C$ required by global continuity using the constants given by local continuity on the finite open cover of $D$.

We show the relationships between H{\"o}lder continuity and other forms of continuity: global 1-H{\"o}lder continuity is Lipschitz continuity and local 1-H{\"o}lder continuity is local Lipschitz; global \(\gamma\)-H{\"o}lder implies uniform continuity and both global and local imply regular continuity.

\subsection{Kolmogorov-Chentsov theorem}
We first prove a generalised version of Markov's inequality:

\begin{lemma}
Let \((\Omega, \mathcal{A},\mu)\) be a measure space, Let X be a real-valued measurable function on \(\mathcal{A}\). Let \(f: \mathbb{R} \to [0, \infty)\) be monotonically increasing on the set \((0, \infty) \cup ran(X)\). Then, for any \(\varepsilon > 0\) where \(f(\varepsilon) > 0\):
\[\mu [X(p) \ge \varepsilon] \le \frac{\int_\Omega f (X(x)) \mu(dx))}{f(\varepsilon)}\]
\end{lemma}

With this we can prove our main result:

 \begin{theorem}
Let X be a stochastic process that takes values in a polish space, with index set \([0..\infty)\). Assume there exist numbers \(\alpha, \beta, C > 0\) such that X satisfies the following property: 
\[E[d(X_t, X_s)^\alpha] \leq C |t - s|^{1+\beta}, s,t \in \mathbb{R}^+\]

Then there is a modification Y of X such that for any \(\gamma \in (0..\beta/\alpha)\), Y has H{\"o}lder-continuous paths of order \(\gamma\).
 \end{theorem}
\begin{proof}
Because distance is measurable, we have that \(d(X_s, X_t)\) is a real-valued random variable. We can then apply Markov's inequality above, with \(f(x) = x^a\), to obtain the following:
\[P[d(X_s, X_t) \le \varepsilon] \le \frac{E[d(X_t, X_s)]^\alpha}{\varepsilon^\alpha}\]

By assumption, it then follows that
\[P[d(X_s, X_t) \le \varepsilon] \le \frac{C|t-s|^{1+\beta}}{\varepsilon^\alpha}\]

It then follows that \(X_s\) converges in probability to \(X_t\). We will construct our modification as a limit of the process as \(s \to t\), and because \(X_s\) converges in probability this will allow us to show that the limit is equal to \(X_t\) with probability 1, and hence a modification.

Fix \(T > 0\). We construct a H{\"o}lder-continuous modification of X restricted to \([0, T]\), and then show that we can extend these modifications to the interval \([0, \infty)\).
For this, we will define a null set which characterises the paths that are not H{\"o}lder-continuous.

Define
\[A_n = \{\text{Max } \{d(X_{2^{-n}(k-1)}, X_{2^{-n}k}) \mid k \in 1..\left\lfloor 2^nT\right\rfloor\} \ge 2 ^ {-\gamma n}\}\]
\[N = \lim_{n \to \infty} A_n\]

\(P(B_n)\) tends to 0 as \(n \to \infty\), which allows us to show that N is a null set. We then fix \(\omega \in \Omega - N\) and show that \(\lambda t. X_t \omega\) is H{\"o}lder continuous on the dyadic rationals. H{\"o}lder continuity implies uniform continuity, and therefore \(\lim_{s\in D \to t} X_s(\omega)\) is defined. Define \[\tilde{X}_t = \text{if } t \in D \text{ then } X_t(\omega) \text{ else } \lim_{s\in D \to t} X_s(\omega)\] We show that our constructed process is H{\"o}lder continuous, and equal to the original one almost everywhere by convergence in probability.

We have now proven that for any T, there is a H{\"o}lder-\(\gamma\)-continuous modification of X on \([0,T]\). Any two modifications on \([0, S]\) and \([0,T]\) are indistinguishable on \([0, \text{min}(S, T)]\) because they are continuous. Therefore, for \(S, T > 0\) we have a null set \(N_{S,T}\) such that \({X^T \neq X^S} \subseteq N_{S,T}\). Then, \(N' = \bigcup_{S, T \in \mathbb{N}} N_{S, T}\) is a null set by countable subadditivity. Define \(\tilde{X} = X\) on all \(\omega \in \Omega - N'\). Then, \(\tilde{X}\) restricted to T is equal to \(X^T\), and hence H{\"older} continuous on \(\Omega - N'\). Furthermore, it is a modification of X, completing our proof.
\end{proof}

\subsection{Constructing Brownian motion}\label{sec:brownian-motion}

We have now developed the mathematical background we need to construct Brownian motion. We define the family of Brownian motion kernels by:
 \[\kappa_t(\omega, -) \triangleq (\delta_\omega \star \mathcal{N}_{0, t})\]
\noindent These are defined as the convolution between the Dirac point measure \(\delta\) and the normal distribution with variance t. We show that this defines a Markov semigroup, and we can therefore we can use it to construct an infinite-dimensional product space. The coordinate process from this product space have the requisite distributions, and we prove that this process satisfies the requirements of the Kolmogorov-Chentsov theorem for all \(\gamma < \frac{1}{2}\), which yields a continuous modification of our process. We have finally constructed a Brownian motion.
This process can be understood as the limit of a random walk as the size of the steps tends to 0. Indeed, future work involves proving Donsker's invariance theorem which states exactly this. It is a ubiquitous and central process in the study of natural phenomena, and its formalisation paves the way for the mechanisation of advanced results in stochastic process theory.

\section{Evaluation}\label{sec:evaluation}
To evaluate our contributions, we pose two research questions.

\paragraph{Question 1: How well do textbook mathematics lend themselves to mechanisation?}
We found that on a number of occasions, the textbook proofs omitted a significant amount of detail, at times skipping proof obligations entirely. This was most common with measurability proofs, which can be quite tedious but were often dismissed by the author. Indeed, most of the work of formalising pen-and-paper mathematics lies in making the implicit explicit, and filling in the gaps between the broad strokes of the proofs.

In particular, one of the major deviations from Klenke's text was the development of the theory of dyadic intervals. In the proof of Kolmogorov-Chentsov, the author assumes ``without loss of generality'' that $T=1$, implicitly arguing that the reasoning for any other value of $T$ would be identical. This kind of reasoning cannot be carried out ad hoc in a proof assistant, and we therefore had to develop a theory that allowed us to deal with arbitrary values of $T$ in the proof, leading us to dyadic intervals.

Our development consists of $\sim$7000 lines of proofs, comprising 365 lemmas and 55 definitions. The formalisation covers about 20 pages of the textbook, including some preliminary material from the early chapters not previously available in Isabelle/HOL. We consider this to be a substantial piece of work; the pen-and-paper proofs are terse, and significant effort was expended in adapting textbook definitions to work with the Isabelle/HOL libraries and type system.

\paragraph{Question 2: How do the proof facilities of Isabelle/HOL support mechanisation of textbook mathematics?}
Isabelle/HOL was expressive enough for all the definitions we wanted to write, but at times we had to make concessions for the type system. The finite kernel product was defined in the textbook as a recursion using the binary product. This would need dependent types to work in Isabelle, so we had to make this construction explicit. Other such occasions included having to restrict the types of measures to be homogeneous when they could in theory each have a different type, although we doubt this will have any impact in practice. The advantage of using simple types is the wealth of automation facilities available; much of the low-level reasoning can be deferred to automated theorem provers via Sledgehammer~\cite{Blanchette2016Hammers}, making the development simpler. Indeed, throughout the development we count almost 700 subgoals proven via Sledgehammer.

Despite this, we often found ourselves decomposing large proof goals by hand. Sledgehammer is not a panacea, and it goes hand-in-hand with Isabelle's other theorem provers, such as the simplifier and the measurability prover. We made use of two main approaches for improving automation within our theories: first, we made use of type definitions when practical, as these leverage the type system to discharge assumptions automatically. Second, theorem annotations give domain-specific knowledge to the automated provers by indicating which theorems can be applied to specific situation, e.g. simplification rules can be unconditionally applied to simplify a proof goal. Despite this, we still had to write routine proofs by hand, in particular for measurability goals involving the product measure \texttt{PiM} as this makes use of functions as dependent products which can be cumbersome to reason about in Isabelle's type system.

\section{Related work}\label{sec:related-work}
Many current approaches to formal verification depend on significant abstractions or approximations to make analysis tractable. For example, when the continuous space and time are abstracted away robots can be modelled as state machines in tools like RoboChart~\cite{miyazawa19}. We can also model them as hybrid systems with tools like KeYmaera X~\cite{keymaerax} or Isabelle/HOL~\cite{foster21}, which combine continuous behaviour as described by ODEs and discrete jumps. However, there is currently no formal verification tool which targets SHSs directly.

Lavaei et al. conducted a survey on the formal verification of stochastic hybrid systems~\cite{lavaei21}. Most approaches involve discretisation of the SHSs in order to make the automated verification tractable, using techniques such as model checking and reachability analysis. Discretising a system reduces the accuracy of the model, in contrast to the deductive verification approach we propose which reasons about the continuous system.

Logics for verification of stochastic hybrid systems have been described, but to the authors' knowledge no implementations exist. Platzer introduces his stochastic differential dynamic logic SdL~\cite{platzer11}, which extends the differential dynamic logic of KeYmaera X with stochastic differential equations and random assignment. He includes inference rules for differential invariants over stochastic hybrid programs. Similarly, Wang et al. introduce Stochastic Hybrid CSP~\cite{wang17} which tackles a similar challenge, but by instead extending hybrid CSP.

The probability theory of Isabelle/HOL was developed during H{\"o}lzl's PhD thesis~\cite{holzlphd}, in which he introduces basic concepts such as the construction of probability spaces as well as product spaces and Markov Chains. Immler formalised the Kolmogorov extension theorem~\cite{immler12}, which proves the existence of stochastic processes with arbitrary indices. We make use of these results, which are all available in the \textsf{HOL-Analysis} and \textsf{HOL-Probability} libraries of Isabelle/HOL.

There are overlaps between our work and other Isabelle/HOL developments. In their formalisation of quasi-Borel spaces, Hirata et al. developed a theory of measure kernels in order to describe probabilistic programs~\cite{hirata23}. Their definition is compatible with ours, but they do not define any form of the product kernel and thus it is not enough to describe stochastic processes. We do however make use of their formalisation of measurable isomorphisms. Stochastic processes have been covered in Isabelle/HOL too: Keskin et al. have developed a theory martingales in Isabelle/HOL~\cite{keskin23}. The theorems they prove are largely disjoint from ours, but again their definition of stochastic processes is compatible with ours.

The approach we took to mechanising Brownian motion is not the only possible one. It can be constructed directly without need for the Kolmogorov-Chentsov theorem, e.g. L{\`e}vy's construction. However, our approach is more general, and makes it possible to develop the theory of continuous-time stochastic processes in Isabelle/HOL much simpler by obviating the need to construct processes ad hoc in this manner.

\section{Conclusion}\label{sec:conclusion}
We have developed a theory of stochastic processes in Isabelle/HOL, based on the existing probability theory library. We formalised stochastic processes and stochastic kernels, defined a way of constructing processes with kernels, and proved the Kolmogorov-Chentsov theorem for producing continuous modifications of well-behaved processes. We applied this theory in order to construct the Brownian motion process.

In summary, we had to make concessions and interpret the text in order to produce our theory, but our proofs generally followed the same steps as the textbook. We have developed a method for constructing stochastic processes with stationary, independent increments which are continuous almost everywhere, and we then applied this method to construct Brownian motion. This is an important step towards developing a theory of stochastic differential equations in Isabelle/HOL, which will enable the deductive verification of stochastic hybrid systems.

We plan to build on this work by developing a practical verification tool for SHS based on the hybrid systems verification tool by Foster et al.\cite{foster21}. Our objective is to be able to reason about safety and liveness properties of SHS, using a formalised mathematical background that gives us confidence in our foundations. With such a tool, we will be able to provide formally verified confidence bounds on the behaviour of a noisy system.

\printbibliography
\end{document}